\documentclass[preprint,superscriptaddress,amsmath,amssymb,aps,prd,longbibliography,floatfix,nofootinbib,11pt]{revtex4-2}
\usepackage{graphicx}   
\usepackage[
colorlinks=true,        
citecolor=blue,         
linkcolor=blue,         
urlcolor=blue          
]{hyperref}             
\usepackage{bm}        
\usepackage{xcolor}    
\usepackage{multirow}
\usepackage{mathrsfs}
\usepackage{amsmath}

\begin{document}

\title{Signatures of Novikov-Thorne accretion disks around rotating black holes in semiclassical gravity with trace anomaly}

\author{Yuxiang Zuo}
\affiliation{School of Mathematics and Physics, University of South China, Hengyang, 421001 People's Republic of China}

\author{Shiyang Hu}
\email{husy\_arcturus@163.com}
\affiliation{School of Mathematics and Physics, University of South China, Hengyang, 421001 People's Republic of China}

\author{Dan Li}
\affiliation{School of Mathematics and Physics, University of South China, Hengyang, 421001 People's Republic of China}

\author{Chen Deng}
\affiliation{School of Astronomy and Space Science, Nanjing University, Nanjing, 210023 People's Republic of China}

\author{Guansheng He}
\affiliation{School of Mathematics and Physics, University of South China, Hengyang, 421001 People's Republic of China}

\author{Wenbin Lin}
\affiliation{School of Mathematics and Physics, University of South China, Hengyang, 421001 People's Republic of China}

\begin{abstract}
Semiclassical gravity provides a promising framework to address the black hole singularity problem through quantum backreaction characterized by a coupling parameter, making it essential to thoroughly explore the accretion properties around such objects. In this paper, within the framework of semiclassical gravity with type-A trace anomaly, we investigate the thermal radiative signatures of Novikov-Thorne accretion disks around rotating black holes across diverse parameter spaces, focusing on the energy flux density, temperature profile, and thermal luminosity. Our findings demonstrate that the introduction of a positive coupling parameter suppresses disk emission, leading to lower peak values of both energy flux density and temperature, as well as lower peak values and corresponding peak frequencies in thermal luminosity compared to their Kerr counterparts in general relativity. In particular, this discrepancy becomes remarkably pronounced when the black hole possesses a high spin. Mechanistically, increasing the coupling parameter reinforces the effective gravitational field strength of the central compact object, which pushes the equatorial stable circular orbits outward and increases the specific energy of the orbiting matter, thereby reducing the radiative efficiency and weakening the overall disk emission. This suppression mechanism offers a potential astrophysical window to distinguish semiclassical gravity from general relativity and to place observational constraints on the coupling parameter via thermal disk spectra.
\end{abstract}
\maketitle
\section{Introduction}
As one of the cornerstones of modern physics, general relativity (GR) has successfully withstood high-precision experimental and observational tests across diverse scales ranging from the Solar System to cosmological regimes. In particular, as the most emblematic prediction of this theory, the existence of black holes has been confirmed in recent years through laser-interferometric gravitational wave detections and direct imaging observations by the Event Horizon Telescope (EHT) \cite{2016PhRvL.116f1102A,2019ApJ...875L...1E,2022ApJ...930L..12E}. Nonetheless, classical GR inevitably predicts the presence of spacetime singularities at the centers of black holes, where the known laws of physics break down. The unavoidable occurrence of such singularities reveals the intrinsic limitations of GR as a classical field theory, thereby driving the theoretical physics community to actively explore modified gravity theories and quantum gravitational effects beyond the classical framework \cite{2008LRR....11....5R,Carlip:2015asa,2012PhR...513....1C,Shankaranarayanan:2022wbx}.

Semiclassical gravity naturally emerged to resolve the central singularity problem and address the limitations of classical general relativity \cite{1989AnPhy.196..296S}. This framework treats the spacetime background as a classical geometry while quantizing the matter fields, feeding the expectation value of the energy-momentum tensor $\langle T_{\mu\nu}\rangle$ induced by quantum vacuum fluctuations back into the Einstein field equations, which generates quantum backreaction effects. In particular, within high-curvature regimes, quantum vacuum fluctuations can generate an effective negative energy and quantum repulsive effects that violate the classical energy conditions, thereby halting complete gravitational collapse and smoothing the central curvature divergence, offering a self-consistent physical avenue to fundamentally resolve the classical singularity problem. Within this theoretical architecture, the trace anomaly of conformal or massless quantum fields plays a central role \cite{1974NCimA..23..173C,1994CQGra..11.1387D,2023PhRvD.108f1502F}. At the classical level, the trace of the energy-momentum tensor for conformally invariant matter vanishes identically. Following quantization and renormalization, however, this classical scale symmetry is broken by quantum vacuum fluctuations, giving rise to a nonvanishing trace completely determined by the geometric invariants of the background spacetime. Specifically, for the Type-A trace anomaly \cite{1993PhLB..309..279D}, the trace of the energy-momentum tensor couples directly to the four-dimensional Gauss-Bonnet topological invariant $E_{4}$, yielding $g^{\mu\nu}\langle T_{\mu\nu}\rangle=\alpha E_{4}/2$, where $\alpha$ denotes the coupling parameter characterizing the strength of quantum effects, and $g^{\mu\nu}$ represents the contravariant metric tensor. Without relying on specific assumptions about microscopic quantum states, this backreaction term provides a transparent and rigorous analytic route to describe non-perturbative quantum geometric corrections in strong gravitational fields.

Within the framework of semiclassical gravity, several authors have derived black hole solutions \cite{Cai:2009ua,Calza:2022szy}. Interestingly, in the presence of a Type-A trace anomaly, Fernandes derived a rotating black hole solution \cite{2023PhRvD.108f1502F}. The primary distinction between this geometry and the classical Kerr metric lies in the replacement of the constant black hole mass $M$ with an effective mass function $\mathcal{M}(r,\theta)$ that explicitly depends on both the radial coordinate $r$ and polar angle $\theta$. Consequently, the event horizon geometry depends directly on the coupling parameter $\alpha$, while the hidden spacetime symmetry is broken due to the absence of the Carter constant. Such an intriguing black hole solution exhibits distinctive geometric properties, serving not only as a theoretical testbed for gravity theories, but also as an indispensable avenue for testing semiclassical gravity against high-precision astrophysical observations. Building upon these considerations, Zhang et al. employed relativistic ray-tracing techniques \cite{Hu:2020usx,Zhong:2021mty,Hou:2022eev} to numerically simulate the critical curves and optical images of rotating black holes with the Type-A trace anomaly \cite{2024ChPhC..48h5106Z}. Their investigation revealed observable differences between this black hole and the Kerr counterpart, identifying potential observational signatures associated with the coupling parameter $\alpha$. It should be emphasized that the current angular resolution of the EHT remains insufficient to place definitive constraints on subtle geometric deviations in black hole shadows \cite{2018NatAs...2..585M}. Therefore, it is both valuable and necessary to explore distinctive signatures of rotating black holes in semiclassical gravity with the Type-A trace anomaly across other astrophysical processes.

The electromagnetic radiative features of matter accreting onto black holes, including the energy flux density, temperature profile, and emission spectrum, naturally encode the geometric information of the background curved spacetime, providing another robust avenue to probe extreme gravitational regimes and constrain spacetime parameters. Within the standard description of geometrically thin and optically thick accretion flows, the Novikov-Thorne model establishes a classic theoretical framework for steady-state relativistic accretion disks \cite{1973A&A....24..337S,Novikov,1974ApJ...191..499P,1974ApJ...191..507T}. Based on thermodynamic and hydrodynamic equilibrium, this model assumes that the accreting material moves along Keplerian circular orbits in the equatorial plane, with the inner edge of the disk terminating at the innermost stable circular orbit of timelike particles. The local thermal radiation and dynamical properties across the disk are completely determined by the equatorial metric components and their radial derivatives, yielding rigorous analytical formulations. Consequently, the Novikov-Thorne disk model has been extensively applied to electromagnetic emission investigations \cite{Torres:2002td,Boshkayev:2023fft,Capozziello:2025ycu,Pun:2008ae,Perez:2012bx,Pun:2008ua,Harko:2009rp,Harko:2009kj,Chen:2011wb,Chen:2011rx,Perez:2017spz,Karimov:2018whx,Jiang:2024njc,Heydari-Fard:2020ugv,Zuluaga:2021vjc,Collodel:2021gxu,Boshkayev:2021chc,Liu:2020vkh,Heydari-Fard:2020iiu,Heydari-Fard:2021ljh,Kazempour:2022asl,He:2022lrc,Fathi:2023ccx,Cimdiker:2023zdi,Alloqulov:2024zln,Wu:2024sng,Heydari-Fard:2023kgf,Dyadina:2023exu,Boshkayev:2023ipb,Liu:2024brf,Shu:2024tut,Faraji:2025aeg,Nozari:2025zlt,Nozari:2025veb,Liu:2025hhg,Tang:2025sbh,Kumar:2025sxj,Boshkayev:2022vlv,Kurmanov:2021uqv,Heydari-Fard:2024wgu,Kobialko:2025sls,Capolupo:2025dxx,Mummery:2024mrq}, parameter constraints \cite{2011ApJ...731..121B,Bambi:2015ldr}, black hole imaging \cite{Luminet:1979nyg,Zhang:2021hit,Zhang:2022klr,Hu:2023bzy,Guo:2023zwy,Huang:2023ilm,Cui:2024wvz,Li:2024kyv,Li:2025ixk,Yin:2025coq,Cai:2025rst,Cai:2025pan,Gyulchev:2026kvp}, and observational tests across various gravitational backgrounds \cite{2005ApJS..157..335L,2011ApJ...742...85G,2014ApJ...790...29G,Moore:2015bxa,Tripathi:2021rqs,Zhang:2021ymo,Zhao:2021bhj}. Motivated by these advantages, we systematically compute the Novikov-Thorne accretion disk properties around rotating black holes characterized by the coupling parameter $\alpha$ within the framework of semiclassical gravity. This work reveals the joint modulation effects of the observation inclination angle, black hole spin, and quantum coupling parameter on thermal radiation signals, thereby establishing an observational basis to test semiclassical gravity from a qualitative perspective. Furthermore, utilizing observational benchmarks for the supermassive black holes in M87 and the Galactic Center, we have mapped out the ranges of both the peak thermal luminosity and its corresponding peak frequency across the explored parameter space for accretion disks around rotating black holes in semiclassical gravity.

The remainder of this paper is organized as follows. In the next section, we introduce the mass function and spacetime metric of the rotating black hole in semiclassical gravity with a Type-A trace anomaly, and we further determine the innermost stable circular orbit, circular orbits, and the corresponding kinematic quantities of timelike particles. In Section III, based on the Novikov-Thorne framework for geometrically thin and optically thick accretion disks, we evaluate the energy flux density, temperature profile, and thermal luminosity of the disk around the target black hole, demonstrating the relations between these physical observables and the spin and coupling parameters. In the final section, we present our conclusions alongside a brief discussion.
\section{Rotating black holes in semiclassical gravity with type-A trace anomaly}
\subsection{Metric}
In the framework of semiclassical gravity with a Type-A trace anomaly, Fernandes derived a rotating black hole solution \cite{2023PhRvD.108f1502F}. The primary difference between this metric and the classical Kerr spacetime lies in replacing the constant ADM mass parameter $M$ with a coordinate-dependent mass function, $M \rightarrow \mathcal{M}\left(r,\theta\right)$, given by
\begin{equation}\label{1}
\mathcal{M}\left(r,\theta\right) = \frac{2M}{1+\sqrt{1-\frac{8\alpha r\xi M}{\Sigma^{3}}}}.
\end{equation}
Here, $\alpha$ denotes the coupling parameter, whose influence on the spacetime mass distribution and whose admissible parameter space were comprehensively examined in \cite{2024ChPhC..48h5106Z}. The auxiliary functions $\xi$ and $\Sigma$ are defined respectively as $\xi=r^{2}-3a^{2}\cos^{2}\theta$ and $\Sigma=r^{2}+a^{2}\cos^{2}\theta$, where $a$ represents the spin parameter of the black hole.

Because our investigation focuses on the radiation from a geometrically thin accretion disk lying in the equatorial plane within the Novikov-Thorne framework, the analysis can be restricted to $\theta=\pi/2$. On the equatorial plane, one has $\xi=\Sigma=r^{2}$, and the effective mass function reduces to $\mathcal{M} _{e}=\mathcal{M}\left(r,\pi/2\right)=2M/\left(1+\sqrt{1-\frac{8\alpha M}{r^{3}}}\right)$. Under this condition, the equatorial metric of the rotating black hole in semiclassical gravity with a Type-A trace anomaly is expressed as
\begin{equation}\label{2}
\textrm{d}s^{2}=-\left(1-\frac{2r\mathcal{M}_{e}}{\Sigma}\right)\textrm{d}t^{2}+\frac{\Sigma}{\Delta}\textrm{d}r^{2}+\left(r^{2}+a^{2}+\frac{2ra^{2}\mathcal{M}_{e}}{\Sigma}\right)\textrm{d}\varphi^{2}-\frac{4ra\mathcal{M}_{e}}{\Sigma}\textrm{d}t\textrm{d}\varphi,
\end{equation}
with $\Delta=r^{2}-2r\mathcal{M}_{e}+a^{2}$.

\begin{figure*}
\centering                   
\includegraphics[width=6cm]{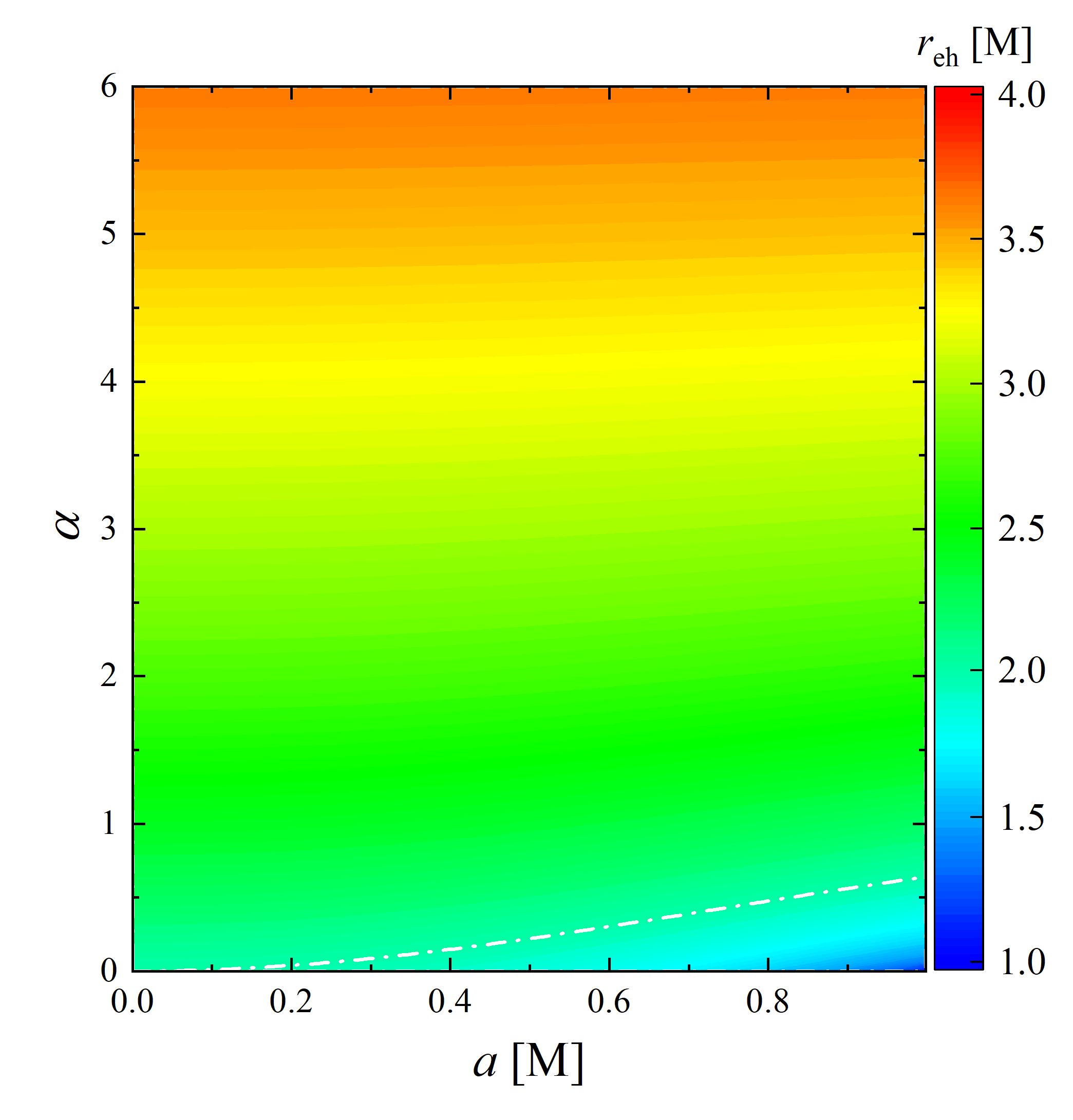}
\caption{Distribution of the event horizon radius $r_{\textrm{eh}}$ in the two-dimensional parameter space spanned by $\alpha$ and $a$. The event horizon radius increases monotonically with the coupling parameter $\alpha$ and decreases with the spin parameter $a$. The white dashed line marks the contour $r_{\textrm{eh}}=2$.}      
\label{fig1}                 
\end{figure*}

When $\Delta=0$, the metric component $\Sigma/\Delta$ diverges, and the corresponding root $r_{\textrm{eh}}$ defines the radius of the event horizon on the equatorial plane. Figure \ref{fig1} illustrates the distribution of the event horizon radius in the two-dimensional parameter space spanned by $\alpha$ and $a$. One can observe that $r_{\textrm{eh}}$ decreases with increasing spin parameter $a$, whereas it grows monotonically with the coupling parameter $\alpha$. Furthermore, a relatively large value of $\alpha$ can overshadow the contraction effect on the horizon induced by an increasing spin parameter. The white dashed line represents the contour of $r_{\textrm{eh}}=2$, which corresponds to the maximum horizon radius predicted by standard GR. Consequently, an event horizon radius exceeding this threshold can serve as a potential observational signature to distinguish semiclassical gravity from standard GR. On the other hand, it is interesting to note that \cite{2024ChPhC..48h5106Z} pointed out that the coupling parameter can take negative values, with the characteristic spacetime quantities, such as the photon ring radius, varying continuously as the coupling parameter transitions from positive to negative. This indicates that our findings can be extended into the negative coupling regime. Nevertheless, when the coupling parameter is negative, only a narrow parameter window allows for the existence of black hole solutions; hence, we maintain the coupling parameter as positive in the subsequent calculations.
\subsection{Effective potential and circular orbit}
Within the Novikov-Thorne framework, the inner edge of the accretion disk coincides with the innermost stable circular orbit (ISCO) of timelike particles in the equatorial plane, while the accreting material outside the ISCO moves along circular orbits. Therefore, it is necessary to first determine the dynamical properties of circular orbits in the target spacetime. According to the equatorial metric \eqref{2}, the Lagrangian $\mathcal{L}$ describing the motion of matter is given by
\begin{equation}\label{3}
2\mathcal{L} = -\left(1-\frac{2r\mathcal{M}_{e}}{\Sigma}\right)\dot{t}^{2}+\frac{\Sigma}{\Delta}\dot{r}^{2}+\left(r^{2}+a^{2}+\frac{2ra^{2}\mathcal{M}_{e}}{\Sigma}\right)\dot{\varphi}^{2}-\frac{4ra\mathcal{M}_{e}}{\Sigma}\dot{t}\dot{\varphi},
\end{equation}
where $\dot{x}=(\dot{t},\dot{r},\dot{\varphi})$ denotes the velocity of the matter, with dots representing differentiation with respect to the proper time $\tau$.

Because the background spacetime does not explicitly depend on the coordinate time $t$ or the azimuthal angle $\varphi$, there exist two constants of motion, namely the specific energy $E$ and the specific angular momentum $L$ of the particle. Their relations to the particle velocity follow from the Euler-Lagrange equations,
\begin{eqnarray}
E = -p_{t} = \left(1-\frac{2r\mathcal{M}_{e}}{\Sigma}\right)\dot{t} + \frac{2ra\mathcal{M}_{e}}{\Sigma}\dot{\varphi}, \label{4} \\
L = p_{\varphi} = \left(r^{2}+a^{2}+\frac{2ra^{2}\mathcal{M}_{e}}{\Sigma}\right)\dot{\varphi} - \frac{2ra\mathcal{M}_{e}}{\Sigma}\dot{t}, \label{5}
\end{eqnarray}
where $p_{\mu}$ represents the canonical momentum of the particle. Expressing $\dot{t}$ and $\dot{\varphi}$ in terms of these constants of motion, we obtain
\begin{eqnarray}
\dot{t} &=& \frac{\mathcal{C}E-\mathcal{B}L}{\Delta}, \label{6} \\
\dot{\varphi} &=& \frac{\mathcal{B}E+\mathcal{A}L}{\Delta}, \label{7}
\end{eqnarray}
with the metric coefficients $\mathcal{A}$, $\mathcal{B}$, and $\mathcal{C}$ given by
\begin{eqnarray}
\mathcal{A} &=& 1-\frac{2\mathcal{M}_{e}}{r}, \label{8} \\
\mathcal{B} &=& \frac{2a\mathcal{M}_{e}}{r}, \label{9} \\
\mathcal{C} &=&  r^{2}+a^{2}+\frac{2a^{2}\mathcal{M}_{e}}{r}. \label{10}
\end{eqnarray}
Substituting $\dot{t}$ and $\dot{\varphi}$ into the Lagrangian \eqref{3} and imposing the timelike normalization condition $2\mathcal{L}=-1$, we have
\begin{equation}\label{11}
\frac{r^{2}}{\Delta}\dot{r}^{2}+\frac{1}{\Delta}\left(\mathcal{A}L^{2}+2\mathcal{B}EL-\mathcal{C}E^{2}\right)=-1.
\end{equation}
Furthermore, we obtain the radial equation of motion for timelike particles,
\begin{equation}\label{12}
\dot{r}^{2}+V_{\textrm{eff}}=E^{2}.
\end{equation}
Here, $V_{\textrm{eff}}$ denotes the effective potential governing timelike particles moving in the equatorial plane of the target spacetime, which can be expressed analytically as
\begin{equation}\label{13}
V_{\textrm{eff}} = 1-\frac{2\mathcal{M}_{e}}{r}+\frac{L^{2}-a^{2}\left(E^{2}-1\right)}{r^{2}}-\frac{2\mathcal{M}_{e}\left(L-aE\right)^{2}}{r^{3}}.
\end{equation}

\begin{figure*}
\centering                   
\includegraphics[width=6cm]{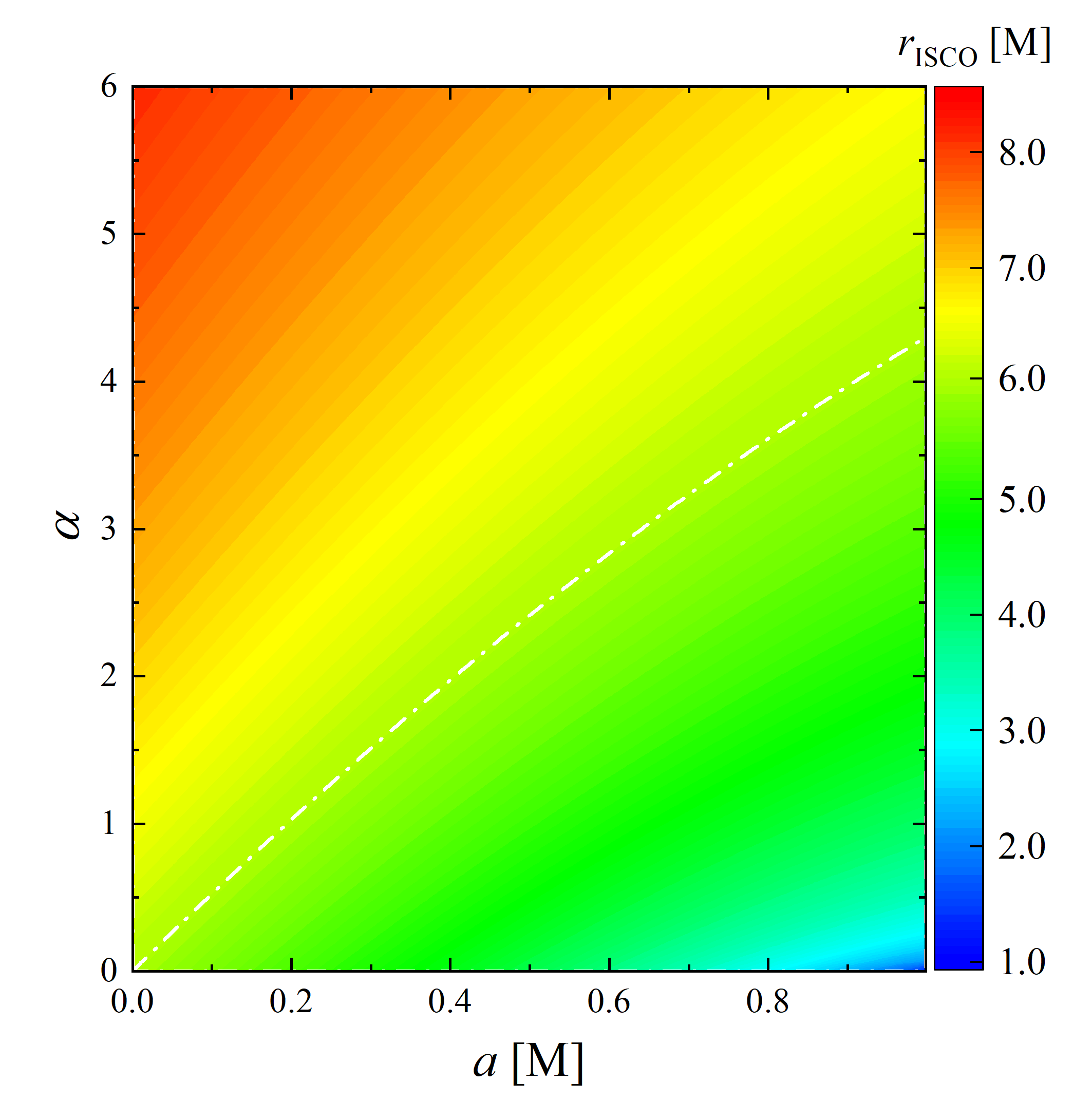}
\caption{Distribution of the ISCO radius $r_{\textrm{ISCO}}$ in the two-dimensional parameter space spanned by $\alpha$ and $a$. The ISCO radius increases monotonically with the coupling parameter $\alpha$ and decreases with the spin parameter $a$. The white dashed line marks the contour $r_{\textrm{ISCO}}=6$, corresponding to the maximum ISCO radius for prograde timelike particles within standard GR.}      
\label{fig2}                 
\end{figure*}

For circular orbits, the radial velocity vanishes, $\dot{r}=0$. Imposing the marginal stability conditions $\partial V_{\textrm{eff}}/\partial r = \partial^{2} V_{\textrm{eff}}/\partial r^{2}=0$, we determine the radius of the ISCO. Figure \ref{fig2} presents the distribution of the ISCO radius $r_{\textrm{ISCO}}$ for equatorial timelike particles in the two-dimensional parameter space spanned by $\alpha$ and $a$, confirming that $r_{\textrm{ISCO}}$ is positively correlated with the coupling parameter $\alpha$. Furthermore, distinct from the behavior observed in Fig. \ref{fig1}, the impact of the spin parameter $a$ on the ISCO radius remains readily discernible even when $\alpha$ takes relatively large values. The white dashed line corresponds to $r_{\textrm{ISCO}}=6$, above which lies the parameter domain inaccessible to prograde timelike particles in the classical Kerr spacetime. From another perspective, a larger ISCO radius typically signals a stronger effective gravitational field of the central object; in other words, introducing the coupling parameter enhances the local gravitational attraction, causing the ISCO to settle at larger radial distances.

\begin{figure*}
\centering                   
\includegraphics[width=16cm]{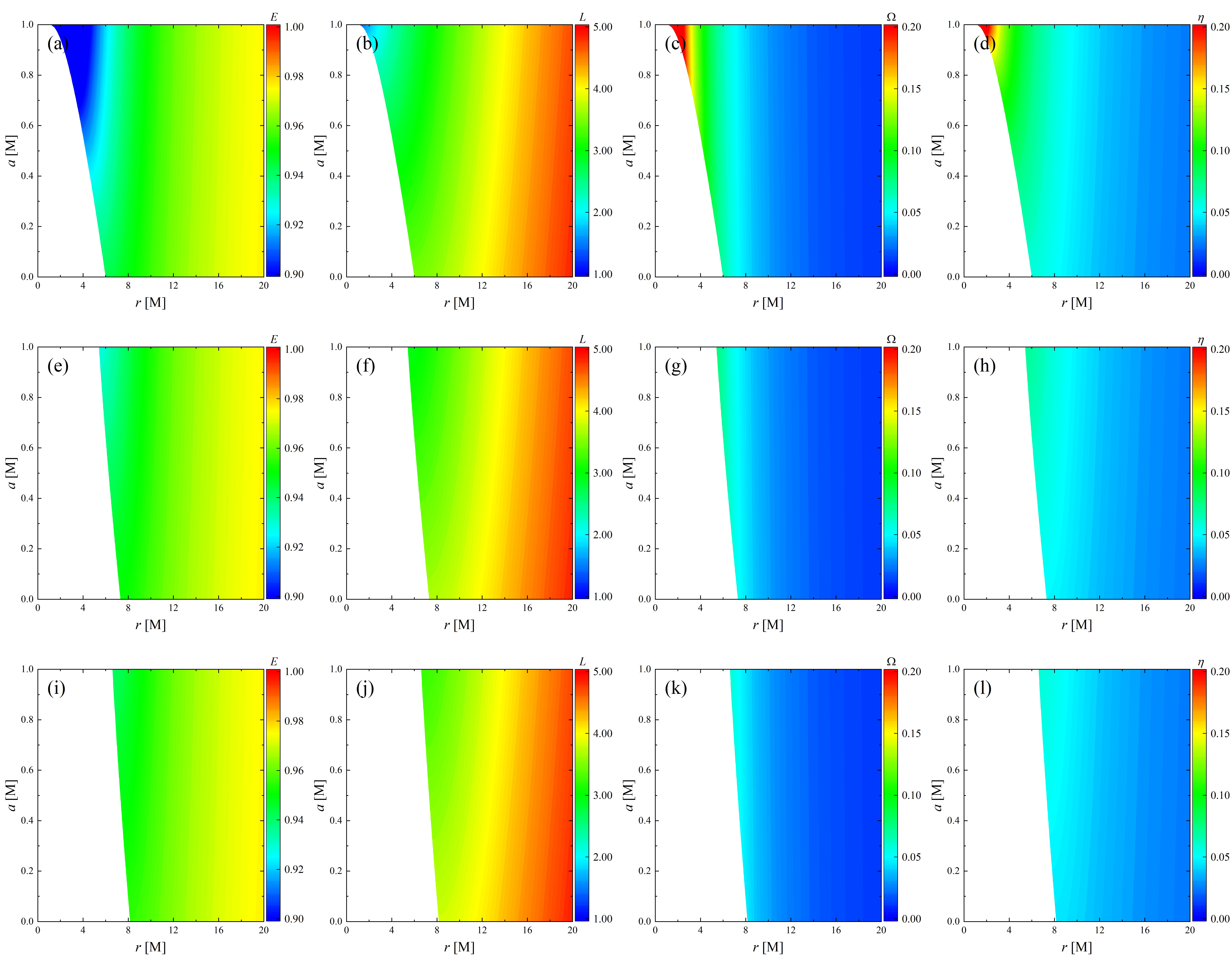}
\caption{Distributions of the specific energy $E$, specific angular momentum $L$, orbital angular velocity $\Omega$, and radiative efficiency $\eta$ for stable circular orbits in the two-dimensional parameter space of $a$ and $r$ under fixed coupling parameters $\alpha$. The rows from top to bottom correspond to coupling parameters of $\alpha=$ $0$, $3$, and $6$ respectively. One can observe that introducing and increasing $\alpha$ leads to higher values of $L$, $\Omega$, and $E$, which consequently reduces $\eta$.}      
\label{fig3}                 
\end{figure*}

Next, choosing an arbitrary radius $r$ within the interval $[r_{\textrm{ISCO}},20]$ and simultaneously solving $V_{\textrm{eff}}=E^{2}$ and $\partial V_{\textrm{eff}}/\partial r =0$, we obtain the analytical expressions for the specific energy $E$, specific angular momentum $L$, and orbital angular velocity $\Omega$ for prograde circular orbits:
\begin{eqnarray}
E &=& \frac{\mathcal{A}+\mathcal{B}\Omega}{\sqrt{\mathcal{A}+2\mathcal{B}\Omega-\mathcal{C}\Omega^{2}}}, \label{14} \\
L &=& \frac{-\mathcal{B}+\mathcal{C}\Omega}{\sqrt{\mathcal{A}+2\mathcal{B}\Omega-\mathcal{C}\Omega^{2}}}, \label{15} \\
\Omega &=& \frac{\mathcal{B}^{\prime}+\sqrt{\left(-\mathcal{B}^{\prime}\right)^{2}-\mathcal{C}^{\prime}\left(-\mathcal{A}^{\prime}\right)}}{\mathcal{C}^{\prime}}, \label{16}
\end{eqnarray}
where primes denote partial derivatives with respect to the radial coordinate $r$. On this basis, we evaluate the specific energy $E$, specific angular momentum $L$, and orbital angular velocity $\Omega$, and radiative efficiency $\eta$ of circular orbits under various coupling parameters, where $\eta=1-E$. The corresponding results are displayed in Fig. \ref{fig3}. One can observe that with an increasing circular orbit radius, both the specific energy and specific angular momentum increase monotonically, whereas the orbital angular velocity and radiative efficiency decrease continuously. This radial trend is precisely opposite to the effects induced by an increasing spin parameter. Moreover, apart from shifting the curves outward because of the expansion of the ISCO, increasing the coupling parameter $\alpha$ slightly enhances the specific energy, specific angular momentum, and angular velocity while reducing the orbital radiative efficiency, although these modifications remain quantitatively modest. For instance, fixing the orbital radius at $r=10$ and the spin parameter at $a=0.91$, as the coupling parameter $\alpha$ increases from $0$ to $6$, the specific energy $E$ grows from $0.95228$ to $0.95409$, the specific angular momentum $L$ rises from $3.45709$ to $3.55934$, and $\Omega$ increases from $0.03068$ to $0.03140$. It is worth emphasizing that the reduction in radiative efficiency caused by an increased coupling parameter indicates a diminished capacity of the accreting material to convert gravitational potential energy into electromagnetic radiation, thereby implying a decrease in the energy flux density, temperature profile, and thermal luminosity of the accretion disk.
\section{Accretion disk features}
Within the Novikov-Thorne accretion disk framework, the disk is assumed to be steady-state, axisymmetric, geometrically thin, and optically thick, satisfying vertical hydrostatic equilibrium and local thermal equilibrium. Angular momentum is transported outward by an effective viscosity, while viscous dissipation is balanced by radiative cooling. The local emission is thermal and blackbody-like, resulting in a multi-temperature blackbody spectrum for the entire disk.

The standard expression for the energy flux density $F(r)$ [erg $\textrm{s}^{-1}$ $\textrm{cm}^{-2}$] follows from the conservation laws of rest mass, energy, and angular momentum, depending on the black hole accretion rate $\dot{M}$, the orbital kinematic variables, and the reduced metric determinant $\mathcal{G}$ in the equatorial plane. A zero-torque boundary condition is imposed at the ISCO, meaning that the viscous torque vanishes at the inner edge. This formulation was originally established in \cite{Novikov,1974ApJ...191..499P,1974ApJ...191..507T} and has been extensively presented in the literature \cite{Torres:2002td,Boshkayev:2023fft,Capozziello:2025ycu,Pun:2008ae,Perez:2012bx,Pun:2008ua,Harko:2009rp,Harko:2009kj,Chen:2011wb,Chen:2011rx,Perez:2017spz,Karimov:2018whx,Jiang:2024njc,Heydari-Fard:2020ugv,Zuluaga:2021vjc,Collodel:2021gxu,Boshkayev:2021chc,Liu:2020vkh,Heydari-Fard:2020iiu,Heydari-Fard:2021ljh,Kazempour:2022asl,He:2022lrc,Fathi:2023ccx,Cimdiker:2023zdi,Alloqulov:2024zln,Wu:2024sng,Heydari-Fard:2023kgf,Dyadina:2023exu,Boshkayev:2023ipb,Liu:2024brf,Shu:2024tut,Faraji:2025aeg,Nozari:2025zlt,Nozari:2025veb,Liu:2025hhg,Tang:2025sbh,Kumar:2025sxj,Boshkayev:2022vlv,Kurmanov:2021uqv,Heydari-Fard:2024wgu,Kobialko:2025sls,Capolupo:2025dxx,Mummery:2024mrq,2011ApJ...731..121B,Bambi:2015ldr,Luminet:1979nyg,Zhang:2021hit,Zhang:2022klr,Hu:2023bzy,Guo:2023zwy,Huang:2023ilm,Cui:2024wvz,Li:2024kyv,Li:2025ixk,Yin:2025coq,Cai:2025rst,Cai:2025pan,Gyulchev:2026kvp}. It is given by
\begin{equation}\label{17}
F(r) = -\frac{\dot{M}}{4\pi\sqrt{-\mathcal{G}}}\frac{\Omega^{\prime}}{\left(E-\Omega L\right)^{2}}\int^{r}_{r_{\textrm{ISCO}}}\left(E-\Omega L\right)L^{\prime}\textrm{d}r,
\end{equation}
where the upper integration limit $r$ denotes the radial coordinate of the emitting ring, corresponding to the radius of the circular orbit. According to the Stefan-Boltzmann law, the effective disk temperature $T(r)$ corresponding to the energy flux density $F(r)$ is determined by
\begin{equation}\label{18}
T(r) = \left[\frac{F(r)}{\sigma}\right]^{\frac{1}{4}},
\end{equation}
where $\sigma$ is the Stefan-Boltzmann constant.
\begin{figure*}
\centering                   
\includegraphics[width=16cm]{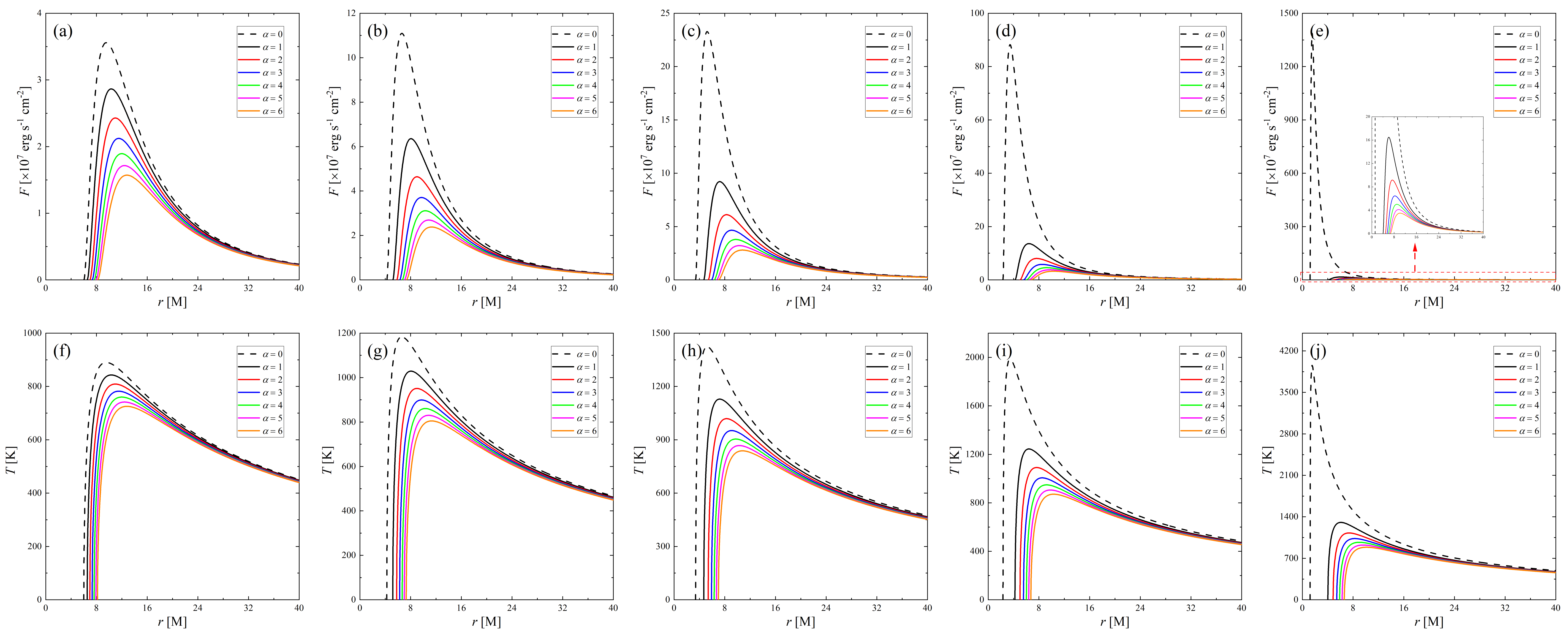}
\caption{Radial profiles of the energy flux density (top row) and the effective temperature (bottom row) of the Novikov-Thorne accretion disk across diverse parameter configurations. From left to right, the spin parameter is set to $a=$ $0$, $0.5$, $0.7$, $0.9$, and $0.9985$, respectively. In each panel, the dashed curve represents the scenario predicted by standard GR. Both $F(r)$ and $T(r)$ decrease monotonically when the coupling parameter $\alpha$ is introduced and enlarged, confirming that the coupling parameter systematically suppresses the disk emission.}      
\label{fig4}                 
\end{figure*}

Adopting a black hole mass of $M = 10^{6}$$M_{\odot}$ and an accretion rate of $\dot{M} = 10^{-12}$$M_{\odot}$$\textrm{yr}^{-1}$, where $M_{\odot}$ denotes the solar mass, we compute the energy flux density and effective temperature profiles of the Novikov-Thorne accretion disk across different values of the coupling parameter $\alpha$ and spin parameter $a$, as illustrated in Fig. \ref{fig4}. The radial behaviors of the energy flux density and temperature exhibit nearly identical evolutionary profiles: both rise sharply from small values at the inner boundary, reach a distinct peak, and subsequently decay monotonically toward larger radii. Moving from left to right across the panels, an increasing spin parameter substantially enhances disk emission, manifested by an elevated peak amplitude and an inward displacement of the peak position. Within each individual panel of the top row, increasing the coupling parameter $\alpha$ suppresses the disk emission while shifting the peak radius outward. This phenomenon arises because matter orbiting closer to the black hole possesses lower specific energy, as evidenced in the leftmost column of Fig. \ref{fig3}, corresponding to a higher radiative efficiency $\eta=1-E$ and thus a boosted radiation flux. The presence of the coupling parameter pushes the accreting material outward, thereby suppressing the overall emission of the disk. Furthermore, the discrepancy in energy flux density between the standard GR Kerr solution and the rotating black hole in semiclassical gravity with a Type-A trace anomaly widens as the spin parameter grows. As displayed in panel (e), the peak flux of the Kerr black hole represented by the dashed curve is nearly $80$ times that of the semiclassical gravity counterpart, suggesting that distinguishing between the two gravity theories becomes much more feasible in the high-spin regime. In light of the monotonic relation prescribed by the Stefan-Boltzmann law, the modulation exerted by $\alpha$ on the temperature profile is qualitatively identical to its effect on the energy flux density.

Based on Fig. \ref{fig4}, it is worth pointing out that for our chosen benchmark mass and accretion rate, the energy flux density and temperature of the Novikov-Thorne disk around a standard Kerr black hole possess theoretical lower thresholds, which are realized when $a=0$ at $r=9.53954$, yielding $F(r) \approx 3.56 \times 10^{7}$ [erg $\textrm{s}^{-1}$ $\textrm{cm}^{-2}$] and $T(r) \approx 890$ [K], respectively. Consequently, if observational measurements infer values lying strictly beneath these baseline thresholds, such detections would provide compelling evidence supporting the breakdown of GR, while simultaneously indicating viable parameter space for semiclassical gravity with a Type-A trace anomaly.

Having acquired the disk temperature distribution, we can further evaluate the thermal luminosity $L$ of the accretion disk \cite{Torres:2002td,Boshkayev:2023fft,Capozziello:2025ycu,Pun:2008ae,Perez:2012bx,Pun:2008ua,Harko:2009rp,Harko:2009kj,Chen:2011wb,Chen:2011rx,Perez:2017spz,Karimov:2018whx,Jiang:2024njc,Heydari-Fard:2020ugv,Zuluaga:2021vjc,Collodel:2021gxu,Boshkayev:2021chc,Liu:2020vkh,Heydari-Fard:2020iiu,Heydari-Fard:2021ljh,Kazempour:2022asl,He:2022lrc,Fathi:2023ccx,Cimdiker:2023zdi,Alloqulov:2024zln,Wu:2024sng,Heydari-Fard:2023kgf,Dyadina:2023exu,Boshkayev:2023ipb,Liu:2024brf,Shu:2024tut,Faraji:2025aeg,Nozari:2025zlt,Nozari:2025veb,Liu:2025hhg,Tang:2025sbh,Kumar:2025sxj,Boshkayev:2022vlv,Kurmanov:2021uqv,Heydari-Fard:2024wgu,Kobialko:2025sls,Capolupo:2025dxx,Mummery:2024mrq}
\begin{equation}\label{19}
L = \frac{8\pi h\cos\gamma}{c^{2}}\int^{r_{\textrm{edge}}}_{r_{\textrm{ISCO}}}\int_{0}^{2\pi}\frac{\nu_{\textrm{e}}^{3}\sqrt{-\mathcal{G}}\textrm{d}r\textrm{d}\varphi}{e^{h\nu_{\textrm{e}}/\left[k_{\textrm{B}} T(r)\right]}-1}.
\end{equation}
Here, $h$ is the Planck constant, $k_{\textrm{B}}$ denotes the Boltzmann constant, and $c$ represents the speed of light. The parameter $\gamma$ denotes the observation inclination angle, defined as the angle between the line of sight and the rotation axis of the black hole. The quantity $r_{\textrm{edge}}$ marks the outer boundary of the accretion disk adopted in our computations, which is set to $r_{\textrm{edge}}=40$ because the radiation from the disk at larger radii is negligible. The variable $\nu_{\textrm{e}}$ is the emission frequency of the accreting matter, related to the observed frequency $\nu$ via $\nu_{\textrm{e}} = g\nu$, where $g$ represents the redshift factor incorporating Doppler effects and gravitational influence on the observed flux, formulated as
\begin{equation}\label{20}
g = \frac{1+\Omega r\sin\varphi\sin\gamma}{\sqrt{\mathcal{A}+2\mathcal{B}\Omega-\mathcal{C}\Omega^{2}}}.
\end{equation}
It is worth pointing out that the analytical relation \eqref{19} for the thermal luminosity does not incorporate light deflection effects, taking into account only gravitational redshift and Doppler boosting. That is, in the regime of high observation inclination angles where gravitational lensing becomes pronounced, Eq. \eqref{19}, which relies solely on direct geometric projection, may omit the contributions from higher-order lensed rays. Nevertheless, from a qualitative perspective, Eq. \eqref{19} remains compelling and robust, particularly under the condition of low observation inclination angles.

\begin{figure*}
\centering                   
\includegraphics[width=16cm]{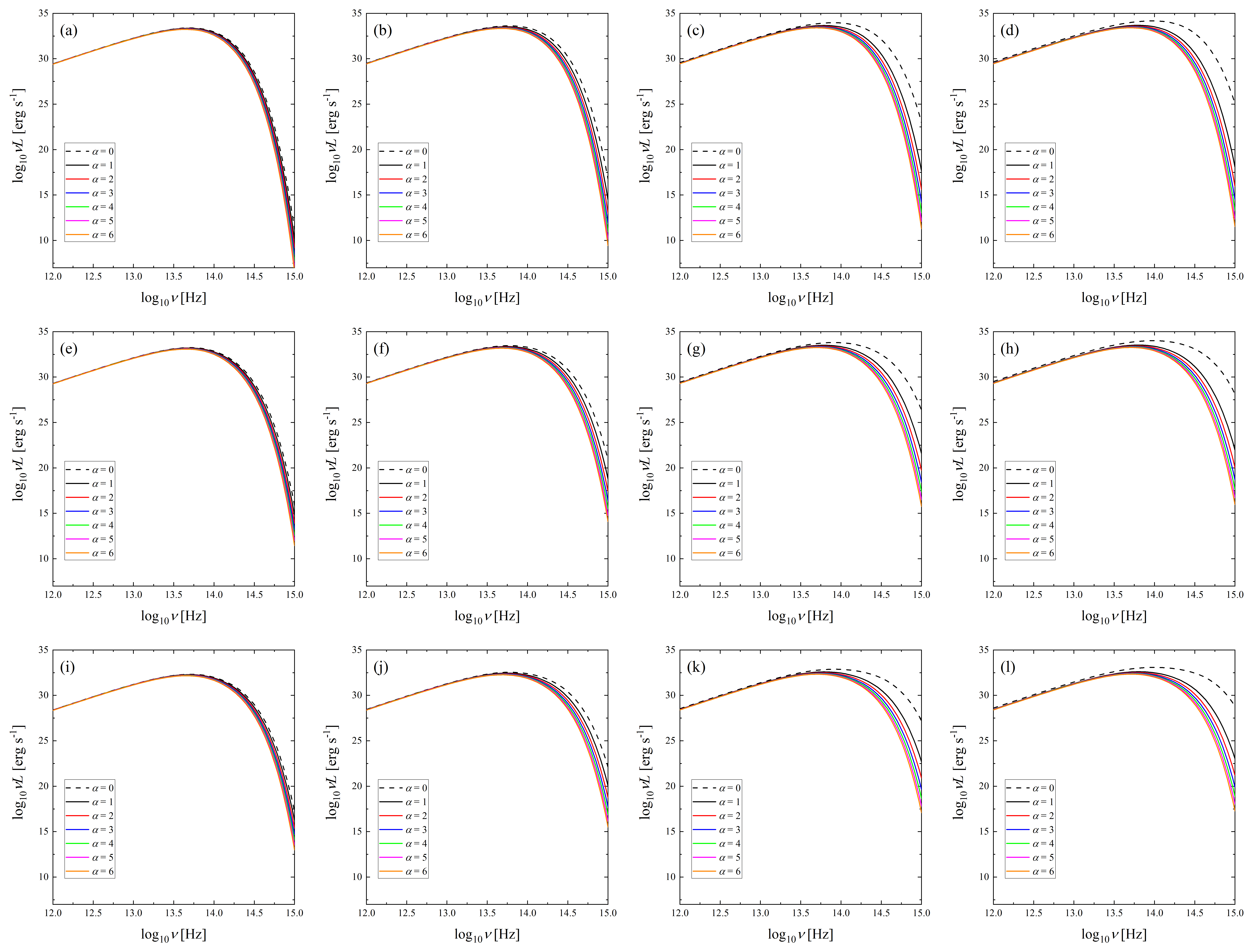}
\caption{Spectral distributions of the accretion disk thermal luminosity as a function of the observed frequency $\nu$ across diverse parameter spaces. From left to right, the spin parameter takes values of $a=$ $0$, $0.54$, $0.94$, and $0.9985$, respectively; from top to bottom, the inclination angles are set to $\gamma=$ $0^{\circ}$, $45^{\circ}$, and $85^{\circ}$, respectively. In each panel, the black dashed curve corresponds to the Kerr black hole prediction. The introduction and increase of the coupling parameter suppresses the thermal luminosity of the disk, reducing both the peak magnitude and the associated peak emission frequency.}      
\label{fig5}                 
\end{figure*}

Adopting the same black hole mass and accretion rate as in Fig. \ref{fig4}, we evaluate the thermal luminosity corresponding to different combinations of the spin parameter $a$, coupling parameter $\alpha$, inclination angle $\gamma$, and observed frequency $\nu$, with the results summarized in Fig. \ref{fig5}. Across all parameter configurations, the thermal luminosity exhibits a universal spectral profile as a function of the observed frequency: it rises steadily at lower frequencies, reaches a maximum, and subsequently declines steeply at higher frequencies. The influence of the remaining three parameters can be characterized as follows: increasing the spin parameter shifts the spectra toward the upper-right domain, producing higher peak luminosities and shifting the peak positions toward higher observed frequencies; conversely, increasing either the observation inclination or the coupling parameter generates the opposite trend, shifting the curves toward the lower-left domain. In addition, an increase in the spin parameter broadens the discrepancy between the dashed and solid curves, indicating that the deviations between rotating black holes in semiclassical gravity with a Type-A trace anomaly and classical Kerr black holes become increasingly prominent. Within each panel, the black dashed curve marks the maximum thermal luminosity from the accretion disk around a neutral rotating black hole predicted by standard general relativity under the specified black hole properties and observational setup. Detecting observational flux values that fall distinctly below this theoretical ceiling would offer supportive evidence for semiclassical gravity and its rotating black hole solution.

Next, we apply the Novikov-Thorne accretion disk model in the framework of semiclassical gravity with a Type-A trace anomaly to M87$^{*}$ and the supermassive black hole at the Galactic Center (Sgr A$^{*}$), examining the thermal luminosity spectra of both sources. In view of the limitations of Eq. \eqref{19} at high inclination angles, the observational constraints on M87$^{*}$, and the fact that the EHT has disfavored inclination angles exceeding $50^{\circ}$ for Sgr A$^{*}$, we set the inclination angle to $\gamma = 17^{\circ}$. Meanwhile, the mass of M87$^{*}$ is adopted as $6.5 \times 10^{9}$ $M_{\odot}$ \cite{2019ApJ...875L...1E}, with its upper and lower bounds on the accretion rate given by $\dot{M}_{1}^{\textrm{max}} = 4.5 \times 10^{-1}$ $M_{\odot}$yr$^{-1}$ and $\dot{M}_{1}^{\textrm{min}} = 4.5 \times 10^{-5}$ $M_{\odot}$yr$^{-1}$ \cite{Drew:2025euq}, respectively. For Sgr A$^{*}$, the mass is taken as $4 \times 10^{6}$ $M_{\odot}$, with the corresponding upper and lower limits on the accretion rate being $\dot{M} _{2}^{\textrm{max}} = 10^{-6}$ $M_{\odot}$yr$^{-1}$ and $\dot{M}_{2}^{\textrm{min}} = 10^{-9}$ $M_{\odot}$yr$^{-1}$ \cite{2022ApJ...930L..12E,EventHorizonTelescope:2022urf}.
\begin{figure*}
\centering                   
\includegraphics[width=16cm]{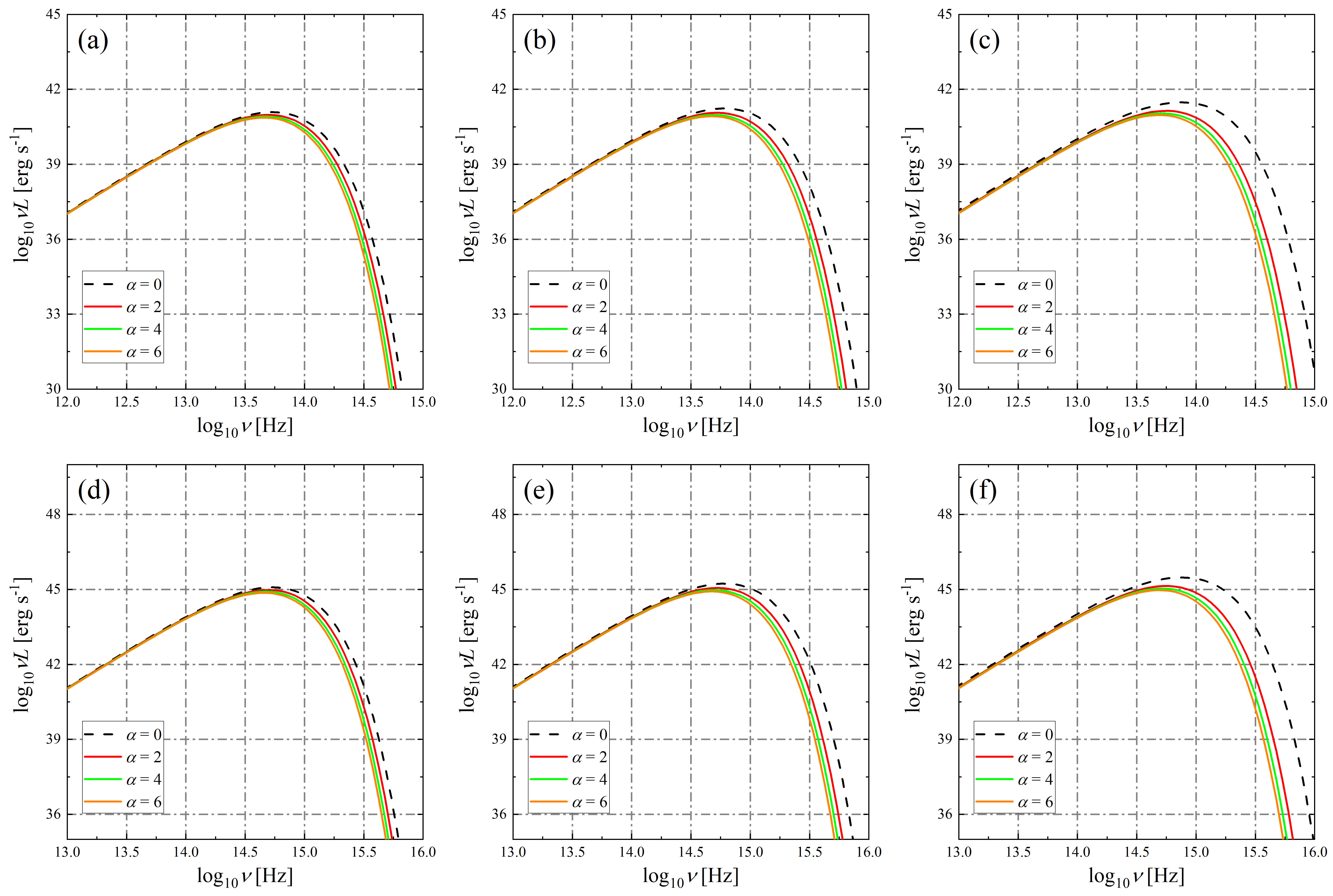}
\caption{Spectral distributions of the accretion disk thermal luminosity as a function of the observed frequency $\nu$ across diverse parameter spaces for M87$^{*}$. From left to right, the spin parameter takes values of $a = 0.3$, $0.6$, and $0.9$, respectively. The top row corresponds to an accretion rate of $\dot{M}_{1}^{\textrm{min}}$, while the bottom row corresponds to $\dot{M}_{1}^{\textrm{max}}$. The black dashed curve represents the Kerr black hole prediction.}      
\label{fig6}                 
\end{figure*}

The thermal luminosity spectra of the accretion disk around M87$^{*}$ corresponding to different accretion rates $\dot{M}_{1}^{\textrm{max},\textrm{min}}$, spin parameters $a$, and coupling parameters $\alpha$ are illustrated in Fig. \ref{fig6}. The influences of the spin and coupling parameters on the spectra are fully consistent with those presented in Fig. \ref{fig5}. In contrast, increasing the accretion rate simultaneously enhances both the peak luminosity of the spectrum and the frequency at which this peak occurs. For example, for the black dashed curve, the peak luminosity in panel (a) is approximately $10^{41}$ [erg s$^{-1}$] at a peak frequency of roughly $10^{13.7}$ [Hz]; these values shift to about $10^{45}$ [erg s$^{-1}$] and $10^{14.7}$ [Hz] in panel (d). Within our explored parameter space, $a \in [0.3,0.9]$ and $\alpha \in [0, 6]$, the peak thermal luminosity and the corresponding frequency range for the black hole accretion disk with a Type-A trace anomaly are $[10^{40.8}, 10^{41.5}]$ [erg s$^{-1}$] and $[10^{13.6},10^{13.9}]$ [Hz] for $\dot{M}_{1}^{\textrm{min}}$, and $[10^{44.9},10^{45.5}]$ [erg s$^{-1}$] and $[10^{14.7},10^{14.9}]$ [Hz] for $\dot{M}_{1}^{\textrm{max}}$, respectively. These intervals can serve as a qualitative, theoretical window for testing semiclassical gravity against astrophysical observations of M87$^{*}$.

\begin{figure*}
\centering                   
\includegraphics[width=16cm]{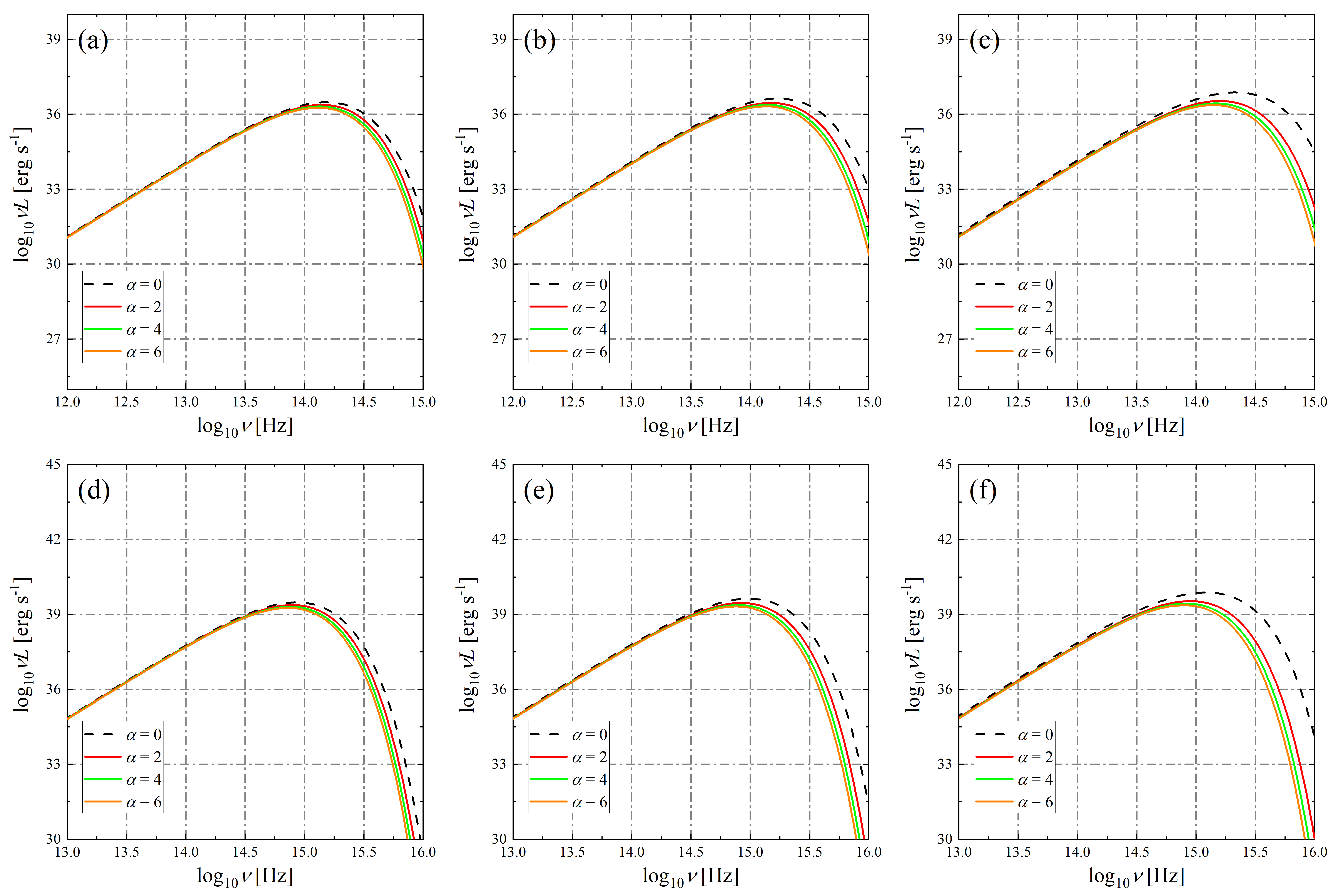}
\caption{Similar to Fig. \ref{fig6}, but for Sgr A$^{*}$, with the top and bottom rows corresponding to the accretion rates $\dot{M}_{2}^{\textrm{min}}$ and $\dot{M}_{2}^{\textrm{max}}$, respectively.}      
\label{fig7}                 
\end{figure*}

Similarly, adopting the mass and accretion rates of Sgr A$^{*}$, we investigate the thermal luminosity of the accretion disk across various parameter configurations, with the results presented in Fig. \ref{fig7}. Within the specified parameter ranges, we calculate that the peak thermal luminosity and its corresponding peak frequency fall within the intervals $[10^{36.2}, 10^{36.9}]$ [erg s$^{-1}$] and $[10^{14.1},10^{14.3}]$ [Hz] for $\dot{M}_{2}^{\textrm{min}}$, and $[10^{39.3},10^{39.9}]$ [erg s$^{-1}$] and $[10^{14.9},10^{15.1}]$ [Hz] for $\dot{M}_{2}^{\textrm{max}}$, respectively.

It is worth pointing out that the computational framework for the Novikov-Thorne accretion disk thermal luminosity spectra involves multiple parameters, such as the black hole spin, mass, accretion rate, observation inclination angle, and quantum coupling parameter. As expected, parameter degeneracies can readily arise. Indeed, from Figs. \ref{fig4}--\ref{fig7}, we can confirm that increasing either the spin parameter or the accretion rate elevates the thermal luminosity spectral profile, whereas increasing the black hole mass or the coupling parameter suppresses disk emission, since $F(r) \sim 1/M^{2}$. Concurrently, increasing the observation inclination angle reduces the effective projected area of the accretion disk, thereby lowering the peak thermal luminosity. These behaviors demonstrate the presence of degeneracies between the spin parameter and the accretion rate, as well as among the black hole mass, coupling parameter, and observation inclination angle in modeling the spectra. Furthermore, even if the inclination angle, black hole mass, and accretion rate are independently constrained through complementary observational avenues, comparable spectral outputs can still be obtained by mutually adjusting the spin and coupling parameters, underscoring an intrinsic degeneracy between the black hole spin and the quantum coupling parameter.
\section{Conclusion and Discussion}
In this paper, we systematically investigated the dynamical and radiative properties of Novikov-Thorne accretion disks around rotating black holes within the framework of semiclassical gravity with a Type-A trace anomaly. We examined the radial profiles of the specific energy, specific angular momentum, orbital angular velocity, and radiative efficiency of the accreting material, evaluated the equatorial distributions of the energy flux density and effective temperature, and analyzed the spectral evolution of the thermal luminosity as a function of the observed frequency, revealing clear radiative distinctions between this semiclassical black hole and the classical Kerr black hole.

Our findings indicate that the inclusion of the quantum coupling parameter reinforces the effective gravitational field of the central black hole, leading to an expansion of both the event horizon and the ISCO. Crucially, the specific energy of stable circular orbits in the equatorial plane increases with the coupling parameter, causing a reduction in the radiative efficiency and suppressing the overall disk emission. Consequently, increasing the coupling parameter lowers the energy flux density, temperature, and thermal luminosity of the Novikov-Thorne disk. This suppression effect becomes increasingly pronounced at higher spin parameters. When system parameters such as the black hole mass, accretion rate, observation inclination angle, and spin are held fixed, the thermal luminosity of the accretion disk around the rotating black hole in semiclassical gravity remains strictly lower than the corresponding Kerr prediction. These systematic radiative signatures offer a theoretical basis for utilizing future high-precision thermal disk observations to test general relativity against semiclassical gravity and to place meaningful constraints on the underlying quantum coupling parameter.

In addition, based on the observed masses and inferred accretion rate intervals for M87$^{*}$ and Sgr A$^{*}$, we computed the corresponding thermal luminosity spectra of the Novikov-Thorne accretion disks. Within our specified parameter space, the peak thermal luminosity and the associated frequency range for M87$^{*}$ are found to span $[10^{40.8},10^{45.5}]$ [erg s$^{-1}$] and $[10^{13.6},10^{14.9}]$ [Hz], respectively, whereas these intervals shift to $[10^{36.2},10^{39.9}]$ [erg s$^{-1}$] and $[10^{14.1},10^{15.1}]$ [Hz] for Sgr A$^{*}$.
\section*{Acknowledgments}
We would like to express our sincere gratitude to the anonymous referee for the valuable comments and suggestions, which have helped us improve this manuscript. This work was supported by the National Natural Science Foundation of China under Grant Nos. 12403081 and 12475057. Guansheng He is partially funded by the Science and Technology Innovation Program of Hunan Province (Grant No. 2026RC3210), the Natural Science Foundation of Hunan Province (Grant No. 2026JJ50351), and the Scientific Research Foundation of the Hunan Provincial Education Department (Grant No. 25B0373).

\bibliographystyle{iopart-num}
\bibliography{references}

@ARTICLE{2016PhRvL.116f1102A,
       author = {{LIGO Scientific Collaboration and Virgo Collaboration}},
        title = "{Observation of Gravitational Waves from a Binary Black Hole Merger}",
      journal = {Phys. Rev. Lett.},
         year = 2016,
        month = feb,
       volume = {116},
       number = {6},
          eid = {061102},
        pages = {061102},
          doi = {10.1103/PhysRevLett.116.061102},
archivePrefix = {arXiv},
       eprint = {1602.03837},
 primaryClass = {gr-qc},
       adsurl = {https://ui.adsabs.harvard.edu/abs/2016PhRvL.116f1102A}
}

@ARTICLE{2019ApJ...875L...1E,
       author = {{Event Horizon Telescope Collaboration}},
        title = "{First M87 Event Horizon Telescope Results. I. The Shadow of the Supermassive Black Hole}",
      journal = {Astrophys. J. Lett.},
         year = 2019,
        month = apr,
       volume = {875},
       number = {1},
          eid = {L1},
        pages = {L1},
          doi = {10.3847/2041-8213/ab0ec7},
archivePrefix = {arXiv},
       eprint = {1906.11238},
 primaryClass = {astro-ph.GA},
       adsurl = {https://ui.adsabs.harvard.edu/abs/2019ApJ...875L...1E}
}

@ARTICLE{2022ApJ...930L..12E,
       author = {{Event Horizon Telescope Collaboration}},
        title = "{First Sagittarius A* Event Horizon Telescope Results. I. The Shadow of the Supermassive Black Hole in the Center of the Milky Way}",
      journal = {Astrophys. J. Lett.},
         year = 2022,
        month = may,
       volume = {930},
       number = {2},
          eid = {L12},
        pages = {L12},
          doi = {10.3847/2041-8213/ac6674},
       adsurl = {https://ui.adsabs.harvard.edu/abs/2022ApJ...930L..12E}
}

@ARTICLE{2008LRR....11....5R,
       author = {{Rovelli}, Carlo},
        title = "{Loop Quantum Gravity}",
      journal = {\lrr},
         year = 2008,
        month = dec,
       volume = {11},
       number = {1},
          eid = {5},
        pages = {5},
          doi = {10.12942/lrr-2008-5},
       adsurl = {https://ui.adsabs.harvard.edu/abs/2008LRR....11....5R}
}

@article{Carlip:2015asa,
    author = "Carlip, Steven and Chiou, Dah-Wei and Ni, Wei-Tou and Woodard, Richard",
    title = "{Quantum Gravity: A Brief History of Ideas and Some Prospects}",
    eprint = "1507.08194",
    archivePrefix = "arXiv",
    primaryClass = "gr-qc",
    doi = "10.1142/S0218271815300281",
    journal = "Int. J. Mod. Phys. D",
    volume = "24",
    number = "11",
    pages = "1530028",
    year = "2015"
}

@ARTICLE{2012PhR...513....1C,
       author = {{Clifton}, Timothy and {Ferreira}, Pedro G. and {Padilla}, Antonio and {Skordis}, Constantinos},
        title = "{Modified gravity and cosmology}",
      journal = {\physrep},
         year = 2012,
        month = mar,
       volume = {513},
       number = {1},
        pages = {1-189},
          doi = {10.1016/j.physrep.2012.01.001},
archivePrefix = {arXiv},
       eprint = {1106.2476},
 primaryClass = {astro-ph.CO},
       adsurl = {https://ui.adsabs.harvard.edu/abs/2012PhR...513....1C}
}

@article{Shankaranarayanan:2022wbx,
    author = "Shankaranarayanan, S. and Johnson, Joseph P.",
    title = "{Modified theories of gravity: Why, how and what?}",
    eprint = "2204.06533",
    archivePrefix = "arXiv",
    primaryClass = "gr-qc",
    doi = "10.1007/s10714-022-02927-2",
    journal = "Gen. Rel. Grav.",
    volume = "54",
    number = "5",
    pages = "44",
    year = "2022"
}

@ARTICLE{1989AnPhy.196..296S,
       author = {{Singh}, T.~P. and {Padmanabhan}, T.},
        title = "{Notes on semiclassical gravity}",
      journal = {Ann. Phys.},
         year = 1989,
        month = dec,
       volume = {196},
       number = {2},
        pages = {296-344},
          doi = {10.1016/0003-4916(89)90180-2},
       adsurl = {https://ui.adsabs.harvard.edu/abs/1989AnPhy.196..296S}
}

@ARTICLE{1974NCimA..23..173C,
       author = {{Capper}, D.~M. and {Duff}, M.~J.},
        title = "{Trace anomalies in dimensional regularization}",
      journal = {Nuovo Cimento A Serie},
         year = 1974,
        month = sep,
       volume = {23},
       number = {1},
        pages = {173-183},
          doi = {10.1007/BF02748300},
       adsurl = {https://ui.adsabs.harvard.edu/abs/1974NCimA..23..173C}
}

@ARTICLE{1994CQGra..11.1387D,
       author = {{Duff}, M.~J.},
        title = "{REVIEW ARTICLE: Twenty years of the Weyl anomaly}",
      journal = {\cqg},
         year = 1994,
        month = jun,
       volume = {11},
       number = {6},
        pages = {1387-1403},
          doi = {10.1088/0264-9381/11/6/004},
archivePrefix = {arXiv},
       eprint = {hep-th/9308075},
 primaryClass = {gr-qc},
       adsurl = {https://ui.adsabs.harvard.edu/abs/1994CQGra..11.1387D}
}

@article{Cai:2009ua,
    author = "Cai, Rong-Gen and Cao, Li-Ming and Ohta, Nobuyoshi",
    title = "{Black Holes in Gravity with Conformal Anomaly and Logarithmic Term in Black Hole Entropy}",
    eprint = "0911.4379",
    archivePrefix = "arXiv",
    primaryClass = "hep-th",
    reportNumber = "KU-TP-039",
    doi = "10.1007/JHEP04(2010)082",
    journal = "\jhep",
    volume = "04",
    pages = "082",
    year = "2010"
}

@article{Calza:2022szy,
    author = "Calz{\'a}, Marco and Casalino, Alessandro and Sebastiani, Lorenzo",
    title = "{Local solutions of general relativity in the presence of the trace anomaly}",
    eprint = "2205.03329",
    archivePrefix = "arXiv",
    primaryClass = "gr-qc",
    doi = "10.1016/j.dark.2022.101066",
    journal = "Phys. Dark Univ.",
    volume = "37",
    pages = "101066",
    year = "2022"
}

@ARTICLE{2023PhRvD.108f1502F,
       author = {{Fernandes}, Pedro G.~S.},
        title = "{Rotating black holes in semiclassical gravity}",
      journal = {\prd},
         year = 2023,
        month = sep,
       volume = {108},
       number = {6},
          eid = {L061502},
        pages = {L061502},
          doi = {10.1103/PhysRevD.108.L061502},
archivePrefix = {arXiv},
       eprint = {2305.10382},
 primaryClass = {gr-qc},
       adsurl = {https://ui.adsabs.harvard.edu/abs/2023PhRvD.108f1502F}
}

@ARTICLE{1993PhLB..309..279D,
       author = {{Deser}, S. and {Schwimmer}, A.},
        title = "{Geometric classification of conformal anomalies in arbitrary dimensions}",
      journal = {\plb},
         year = 1993,
        month = jul,
       volume = {309},
       number = {3-4},
        pages = {279-284},
          doi = {10.1016/0370-2693(93)90934-A},
archivePrefix = {arXiv},
       eprint = {hep-th/9302047},
 primaryClass = {hep-th},
       adsurl = {https://ui.adsabs.harvard.edu/abs/1993PhLB..309..279D}
}

@article{Hu:2020usx,
    author = "Hu, Zezhou and Zhong, Zhen and Li, Peng-Cheng and Guo, Minyong and Chen, Bin",
    title = "{QED effect on a black hole shadow}",
    eprint = "2012.07022",
    archivePrefix = "arXiv",
    primaryClass = "gr-qc",
    doi = "10.1103/PhysRevD.103.044057",
    journal = "Phys. Rev. D",
    volume = "103",
    number = "4",
    pages = "044057",
    year = "2021"
}

@article{Zhong:2021mty,
    author = "Zhong, Zhen and Hu, Zezhou and Yan, Haopeng and Guo, Minyong and Chen, Bin",
    title = "{QED effects on Kerr black hole shadows immersed in uniform magnetic fields}",
    eprint = "2108.06140",
    archivePrefix = "arXiv",
    primaryClass = "gr-qc",
    doi = "10.1103/PhysRevD.104.104028",
    journal = "Phys. Rev. D",
    volume = "104",
    number = "10",
    pages = "104028",
    year = "2021"
}

@article{Hou:2022eev,
    author = "Hou, Yehui and Zhang, Zhenyu and Yan, Haopeng and Guo, Minyong and Chen, Bin",
    title = "{Image of a Kerr-Melvin black hole with a thin accretion disk}",
    eprint = "2206.13744",
    archivePrefix = "arXiv",
    primaryClass = "gr-qc",
    doi = "10.1103/PhysRevD.106.064058",
    journal = "Phys. Rev. D",
    volume = "106",
    number = "6",
    pages = "064058",
    year = "2022"
}

@ARTICLE{2024ChPhC..48h5106Z,
       author = {{Zhang}, Zhenyu and {Hou}, Yehui and {Guo}, Minyong},
        title = "{Observational signatures of rotating black holes in the semiclassical gravity with trace anomaly}",
      journal = {\cpcp},
         year = 2024,
        month = aug,
       volume = {48},
       number = {8},
          eid = {085106},
        pages = {085106},
          doi = {10.1088/1674-1137/ad432b},
archivePrefix = {arXiv},
       eprint = {2305.14924},
 primaryClass = {gr-qc},
       adsurl = {https://ui.adsabs.harvard.edu/abs/2024ChPhC..48h5106Z}
}

@ARTICLE{2018NatAs...2..585M,
       author = {{Mizuno}, Yosuke and {Younsi}, Ziri and {Fromm}, Christian M. and {Porth}, Oliver and {De Laurentis}, Mariafelicia and {Olivares}, Hector and {Falcke}, Heino and {Kramer}, Michael and {Rezzolla}, Luciano},
        title = "{The current ability to test theories of gravity with black hole shadows}",
      journal = {Nat. Astron.},
         year = 2018,
        month = apr,
       volume = {2},
        pages = {585-590},
          doi = {10.1038/s41550-018-0449-5},
archivePrefix = {arXiv},
       eprint = {1804.05812},
 primaryClass = {astro-ph.GA},
       adsurl = {https://ui.adsabs.harvard.edu/abs/2018NatAs...2..585M}
}

@ARTICLE{1973A&A....24..337S,
       author = {{Shakura}, N.~I. and {Sunyaev}, R.~A.},
        title = "{Black holes in binary systems. Observational appearance.}",
      journal = {\aap},
         year = 1973,
        month = jan,
       volume = {24},
        pages = {337-355},
       adsurl = {https://ui.adsabs.harvard.edu/abs/1973A&A....24..337S}
}

@ARTICLE{Novikov,
       author = {{Novikov}, I.~D. and {Thorne}, K.~S.},
        title = "{Astrophysics of black holes}",
      journal = {Black Holes, ed. C. DeWitt and B. DeWitt},
         year = 1973,
        pages = {343}
}

@ARTICLE{1974ApJ...191..499P,
       author = {{Page}, Don N. and {Thorne}, Kip S.},
        title = "{Disk-Accretion onto a Black Hole. Time-Averaged Structure of Accretion Disk}",
      journal = {\apj},
         year = 1974,
        month = jul,
       volume = {191},
        pages = {499-506},
          doi = {10.1086/152990},
       adsurl = {https://ui.adsabs.harvard.edu/abs/1974ApJ...191..499P}
}

@ARTICLE{1974ApJ...191..507T,
       author = {{Thorne}, Kip S.},
        title = "{Disk-Accretion onto a Black Hole. II. Evolution of the Hole}",
      journal = {\apj},
         year = 1974,
        month = jul,
       volume = {191},
        pages = {507-520},
          doi = {10.1086/152991},
       adsurl = {https://ui.adsabs.harvard.edu/abs/1974ApJ...191..507T}
}

@article{Torres:2002td,
    author = "Torres, Diego F.",
    title = "{Accretion disc onto a static nonbaryonic compact object}",
    eprint = "hep-ph/0201154",
    archivePrefix = "arXiv",
    doi = "10.1016/S0550-3213(02)00038-X",
    journal = "Nucl. Phys. B",
    volume = "626",
    pages = "377--394",
    year = "2002"
}

@article{Boshkayev:2023fft,
    author = "Boshkayev, Kuantay and Konysbayev, Talgar and Kurmanov, Yergali and Luongo, Orlando and Muccino, Marco and Taukenova, Aliya and Urazalina, Ainur",
    title = "{Luminosity of accretion disks around rotating regular black holes}",
    eprint = "2307.15003",
    archivePrefix = "arXiv",
    primaryClass = "gr-qc",
    doi = "10.1140/epjc/s10052-024-12446-w",
    journal = "Eur. Phys. J. C",
    volume = "84",
    number = "3",
    pages = "230",
    year = "2024"
}

@article{Capozziello:2025ycu,
    author = "Capozziello, Salvatore and Gambino, Serena and Luongo, Orlando",
    title = "{Comparing Bondi and Novikov{\textendash}Thorne accretion disk luminosity around regular black holes}",
    eprint = "2503.21987",
    archivePrefix = "arXiv",
    primaryClass = "gr-qc",
    doi = "10.1016/j.dark.2025.101950",
    journal = "Phys. Dark Univ.",
    volume = "48",
    pages = "101950",
    year = "2025"
}

@article{Pun:2008ae,
    author = "Pun, C. S. J. and Kovacs, Z. and Harko, T.",
    title = "{Thin accretion disks in f(R) modified gravity models}",
    eprint = "0806.0679",
    archivePrefix = "arXiv",
    primaryClass = "gr-qc",
    doi = "10.1103/PhysRevD.78.024043",
    journal = "Phys. Rev. D",
    volume = "78",
    pages = "024043",
    year = "2008"
}

@article{Perez:2012bx,
    author = "Perez, Daniela and Romero, Gustavo E. and Bergliaffa, Santiago E. Perez",
    title = "{Accretion disks around black holes in modified strong gravity}",
    eprint = "1212.2640",
    archivePrefix = "arXiv",
    primaryClass = "astro-ph.CO",
    doi = "10.1051/0004-6361/201220378",
    journal = "Astron. Astrophys.",
    volume = "551",
    pages = "A4",
    year = "2013"
}

@article{Pun:2008ua,
    author = "Pun, C. S. J. and Kovacs, Z. and Harko, T.",
    title = "{Thin accretion disks onto brane world black holes}",
    eprint = "0809.1284",
    archivePrefix = "arXiv",
    primaryClass = "gr-qc",
    doi = "10.1103/PhysRevD.78.084015",
    journal = "Phys. Rev. D",
    volume = "78",
    pages = "084015",
    year = "2008"
}

@article{Harko:2009rp,
    author = "Harko, Tiberiu and Kovacs, Zoltan and Lobo, Francisco S. N.",
    title = "{Testing Ho{\v{r}}ava-Lifshitz gravity using thin accretion disk properties}",
    eprint = "0907.1449",
    archivePrefix = "arXiv",
    primaryClass = "gr-qc",
    doi = "10.1103/PhysRevD.80.044021",
    journal = "Phys. Rev. D",
    volume = "80",
    pages = "044021",
    year = "2009"
}

@article{Harko:2009kj,
    author = "Harko, Tiberiu and Kovacs, Zoltan and Lobo, Francisco S. N.",
    title = "{Thin accretion disk signatures in dynamical Chern-Simons modified gravity}",
    eprint = "0909.1267",
    archivePrefix = "arXiv",
    primaryClass = "gr-qc",
    doi = "10.1088/0264-9381/27/10/105010",
    journal = "Class. Quant. Grav.",
    volume = "27",
    pages = "105010",
    year = "2010"
}

@article{Chen:2011wb,
    author = "Chen, Songbai and Jing, Jiliang",
    title = "{Thin accretion disk around a Kaluza{\textendash}Klein black hole with squashed horizons}",
    eprint = "1106.5183",
    archivePrefix = "arXiv",
    primaryClass = "gr-qc",
    doi = "10.1016/j.physletb.2011.09.071",
    journal = "Phys. Lett. B",
    volume = "704",
    pages = "641--645",
    year = "2011"
}

@article{Chen:2011rx,
    author = "Chen, Songbai and Jing, Jiliang",
    title = "{Properties of a thin accretion disk around a rotating non-Kerr black hole}",
    eprint = "1110.3462",
    archivePrefix = "arXiv",
    primaryClass = "gr-qc",
    doi = "10.1016/j.physletb.2012.03.047",
    journal = "Phys. Lett. B",
    volume = "711",
    pages = "81--87",
    year = "2012"
}

@article{Perez:2017spz,
    author = "P{\'e}rez, Daniela and Lopez Armengol, Federico G. and Romero, Gustavo E.",
    title = "{Accretion disks around black holes in Scalar-Tensor-Vector Gravity}",
    eprint = "1705.02713",
    archivePrefix = "arXiv",
    primaryClass = "astro-ph.HE",
    doi = "10.1103/PhysRevD.95.104047",
    journal = "Phys. Rev. D",
    volume = "95",
    number = "10",
    pages = "104047",
    year = "2017"
}

@article{Karimov:2018whx,
    author = "Karimov, R. Kh. and Izmailov, R. N. and Bhattacharya, Amrita and Nandi, K. K.",
    title = "{Accretion disks around the Gibbons{\textendash}Maeda{\textendash}Garfinkle{\textendash}Horowitz{\textendash}Strominger charged black holes}",
    eprint = "2002.00589",
    archivePrefix = "arXiv",
    primaryClass = "gr-qc",
    doi = "10.1140/epjc/s10052-018-6270-6",
    journal = "Eur. Phys. J. C",
    volume = "78",
    number = "9",
    pages = "788",
    year = "2018"
}

@article{Jiang:2024njc,
    author = "Jiang, Yu-Hao and Wang, Towe",
    title = "{Accretion disks around magnetically charged black holes in string theory with an Euler-Heisenberg correction}",
    eprint = "2408.10150",
    archivePrefix = "arXiv",
    primaryClass = "gr-qc",
    doi = "10.1103/PhysRevD.110.103009",
    journal = "Phys. Rev. D",
    volume = "110",
    number = "10",
    pages = "103009",
    year = "2024"
}

@article{Heydari-Fard:2020ugv,
    author = "Heydari-Fard, Mohaddese and Heydari-Fard, Malihe and Sepangi, Hamid Reza",
    title = "{Thin accretion disks and charged rotating dilaton black holes}",
    eprint = "2004.05552",
    archivePrefix = "arXiv",
    primaryClass = "gr-qc",
    doi = "10.1140/epjc/s10052-020-7911-0",
    journal = "Eur. Phys. J. C",
    volume = "80",
    number = "4",
    pages = "351",
    year = "2020"
}

@article{Zuluaga:2021vjc,
    author = "Zuluaga, Fabi{\'a}n H. and S{\'a}nchez, Luis A.",
    title = "{Accretion disk around a Schwarzschild black hole in asymptotic safety}",
    eprint = "2106.03140",
    archivePrefix = "arXiv",
    primaryClass = "gr-qc",
    doi = "10.1140/epjc/s10052-021-09644-1",
    journal = "Eur. Phys. J. C",
    volume = "81",
    number = "9",
    pages = "840",
    year = "2021"
}

@article{Collodel:2021gxu,
    author = "Collodel, Lucas G. and Doneva, Daniela D. and Yazadjiev, Stoytcho S.",
    title = "{Circular Orbit Structure and Thin Accretion Disks around Kerr Black Holes with Scalar Hair}",
    eprint = "2101.05073",
    archivePrefix = "arXiv",
    primaryClass = "astro-ph.HE",
    doi = "10.3847/1538-4357/abe305",
    journal = "Astrophys. J.",
    volume = "910",
    number = "1",
    pages = "52",
    year = "2021"
}

@article{Boshkayev:2021chc,
    author = "Boshkayev, Kuantay and Konysbayev, Talgar and Kurmanov, Ergali and Luongo, Orlando and Malafarina, Daniele and Quevedo, Hernando",
    title = "{Luminosity of accretion disks in compact objects with a quadrupole}",
    eprint = "2106.04932",
    archivePrefix = "arXiv",
    primaryClass = "gr-qc",
    doi = "10.1103/PhysRevD.104.084009",
    journal = "Phys. Rev. D",
    volume = "104",
    number = "8",
    pages = "084009",
    year = "2021"
}

@article{Liu:2020vkh,
    author = "Liu, Cheng and Zhu, Tao and Wu, Qiang",
    title = "{Thin Accretion Disk around a four-dimensional Einstein-Gauss-Bonnet Black Hole}",
    eprint = "2004.01662",
    archivePrefix = "arXiv",
    primaryClass = "gr-qc",
    doi = "10.1088/1674-1137/abc16c",
    journal = "Chin. Phys. C",
    volume = "45",
    number = "1",
    pages = "015105",
    year = "2021"
}

@article{Heydari-Fard:2020iiu,
    author = "Heydari-Fard, Mohaddese and Sepangi, Hamid Reza",
    title = "{Thin accretion disk signatures of scalarized black holes in Einstein-scalar-Gauss-Bonnet gravity}",
    eprint = "2009.13748",
    archivePrefix = "arXiv",
    primaryClass = "gr-qc",
    doi = "10.1016/j.physletb.2021.136276",
    journal = "Phys. Lett. B",
    volume = "816",
    pages = "136276",
    year = "2021"
}

@article{Heydari-Fard:2021ljh,
    author = "Heydari-Fard, Mohaddese and Heydari-Fard, Malihe and Sepangi, Hamid Reza",
    title = "{Thin accretion disks around rotating black holes in 4$D$ Einstein{\textendash}Gauss{\textendash}Bonnet gravity}",
    eprint = "2105.09192",
    archivePrefix = "arXiv",
    primaryClass = "gr-qc",
    doi = "10.1140/epjc/s10052-021-09266-7",
    journal = "Eur. Phys. J. C",
    volume = "81",
    number = "5",
    pages = "473",
    year = "2021"
}

@article{Kazempour:2022asl,
    author = "Kazempour, Sobhan and Zou, Yuan-Chuan and Akbarieh, Amin Rezaei",
    title = "{Analysis of accretion disk around a black hole in dRGT massive gravity}",
    eprint = "2203.05190",
    archivePrefix = "arXiv",
    primaryClass = "gr-qc",
    doi = "10.1140/epjc/s10052-022-10153-y",
    journal = "Eur. Phys. J. C",
    volume = "82",
    number = "3",
    pages = "190",
    year = "2022"
}

@article{He:2022lrc,
    author = "He, Tong-Yu and Cai, Ziqiang and Yang, Rong-Jia",
    title = "{Thin accretion disks around a black hole in Einstein-Aether-scalar theory}",
    eprint = "2208.03723",
    archivePrefix = "arXiv",
    primaryClass = "gr-qc",
    doi = "10.1140/epjc/s10052-022-11037-x",
    journal = "Eur. Phys. J. C",
    volume = "82",
    number = "11",
    pages = "1067",
    year = "2022"
}

@article{Fathi:2023ccx,
    author = "Fathi, Mohsen and Cruz, Norman",
    title = "{Observational signatures of a static f(R) black hole with thin accretion disk}",
    eprint = "2304.02111",
    archivePrefix = "arXiv",
    primaryClass = "gr-qc",
    doi = "10.1140/epjc/s10052-023-12341-w",
    journal = "Eur. Phys. J. C",
    volume = "83",
    number = "12",
    pages = "1160",
    year = "2023"
}

@article{Cimdiker:2023zdi,
    author = {{\c{C}}imdiker, {\.I}lim {\.I}rfan and {\"O}vg{\"u}n, Ali and Demir, Durmu{\c{s}}},
    title = "{Thin accretion disk images of the black hole in symmergent gravity}",
    eprint = "2308.03947",
    archivePrefix = "arXiv",
    primaryClass = "gr-qc",
    doi = "10.1088/1361-6382/aceb45",
    journal = "Class. Quant. Grav.",
    volume = "40",
    number = "18",
    pages = "184001",
    year = "2023"
}

@article{Alloqulov:2024zln,
    author = "Alloqulov, Mirzabek and Shaymatov, Sanjar and Ahmedov, Bobomurat and Jawad, Abdul",
    title = "{Radiation properties of the accretion disk around a black hole in Einstein-Maxwell-scalar theory*}",
    eprint = "2401.03184",
    archivePrefix = "arXiv",
    primaryClass = "gr-qc",
    doi = "10.1088/1674-1137/ad137f",
    journal = "Chin. Phys. C",
    volume = "48",
    number = "2",
    pages = "025101",
    year = "2024"
}

@article{Wu:2024sng,
    author = "Wu, Yingdong and Feng, Haiyuan and Chen, Wei-Qiang",
    title = "{Thin accretion disk around black hole in Einstein{\textendash}Maxwell-scalar theory}",
    eprint = "2410.14113",
    archivePrefix = "arXiv",
    primaryClass = "gr-qc",
    doi = "10.1140/epjc/s10052-024-13454-6",
    journal = "Eur. Phys. J. C",
    volume = "84",
    number = "10",
    pages = "1075",
    year = "2024"
}

@article{Heydari-Fard:2023kgf,
    author = "Heydari-Fard, Mohaddese and Heydari-Fard, Malihe and Riazi, Nematollah",
    title = "{Thin accretion disk images of rotating hairy Horndeski black holes}",
    eprint = "2311.12393",
    archivePrefix = "arXiv",
    primaryClass = "gr-qc",
    doi = "10.1007/s10509-024-04359-7",
    journal = "Astrophys. Space Sci.",
    volume = "369",
    number = "9",
    pages = "96",
    year = "2024"
}

@article{Dyadina:2023exu,
    author = "Dyadina, Polina and Avdeev, Nikita",
    title = "{Thin accretion disk signatures in hybrid metric-Palatini gravity}",
    eprint = "2308.02375",
    archivePrefix = "arXiv",
    primaryClass = "gr-qc",
    doi = "10.1140/epjc/s10052-024-12465-7",
    journal = "Eur. Phys. J. C",
    volume = "84",
    number = "1",
    pages = "103",
    year = "2024"
}

@article{Boshkayev:2023ipb,
    author = "Boshkayev, Kuantay and Konysbayev, Talgar and Kurmanov, Yergali and Luongo, Orlando and Muccino, Marco and Quevedo, Hernando and Urazalina, Ainur",
    title = "{Accretion disk in the Hartle{\textendash}Thorne spacetime}",
    eprint = "2306.15050",
    archivePrefix = "arXiv",
    primaryClass = "gr-qc",
    doi = "10.1140/epjp/s13360-024-05072-8",
    journal = "Eur. Phys. J. Plus",
    volume = "139",
    number = "3",
    pages = "273",
    year = "2024"
}

@article{Liu:2024brf,
    author = "Liu, Ailin and He, Tong-Yu and Liu, Ming and Han, Zhan-Wen and Yang, Rong-Jia",
    title = "{Possible signatures of higher dimension in thin accretion disk around brane world black hole}",
    eprint = "2404.14131",
    archivePrefix = "arXiv",
    primaryClass = "gr-qc",
    doi = "10.1088/1475-7516/2024/07/062",
    journal = "\jcap",
    volume = "07",
    pages = "062",
    year = "2024"
}

@article{Shu:2024tut,
    author = "Shu, Yu-Heng and Huang, Jia-Hui",
    title = "{Circular orbits and thin accretion disk around a quantum corrected black hole}",
    eprint = "2412.05670",
    archivePrefix = "arXiv",
    primaryClass = "gr-qc",
    doi = "10.1016/j.physletb.2025.139411",
    journal = "Phys. Lett. B",
    volume = "864",
    pages = "139411",
    year = "2025"
}

@article{Faraji:2025aeg,
    author = "Faraji, Shokoufe",
    title = "{The impact of compact object deformation on thin accretion disk properties}",
    eprint = "2505.17924",
    archivePrefix = "arXiv",
    primaryClass = "gr-qc",
    doi = "10.1140/epjc/s10052-025-13794-x",
    journal = "Eur. Phys. J. C",
    volume = "85",
    number = "2",
    pages = "148",
    year = "2025"
}

@article{Nozari:2025zlt,
    author = "Nozari, Kourosh and Saghafi, Sara and Aliyan, Fatemeh",
    title = "{Investigating QED effects on the thin accretion disk properties around rotating Euler{\textendash}Heisenberg black holes}",
    eprint = "2506.17986",
    archivePrefix = "arXiv",
    primaryClass = "gr-qc",
    doi = "10.1140/epjc/s10052-025-14433-1",
    journal = "Eur. Phys. J. C",
    volume = "85",
    number = "7",
    pages = "735",
    year = "2025"
}

@article{Nozari:2025veb,
    author = "Nozari, Kourosh and Saghafi, Sara and Hajebrahimi, Milad and Jusufi, Kimet",
    title = "{Circular orbits and accretion disk around a deformed-Schwarzschild black hole in loop quantum gravity}",
    eprint = "2504.12486",
    archivePrefix = "arXiv",
    primaryClass = "gr-qc",
    doi = "10.1016/j.dark.2025.102027",
    journal = "Phys. Dark Univ.",
    volume = "49",
    pages = "102027",
    year = "2025"
}

@article{Liu:2025hhg,
    author = "Liu, Shou-Qi and Huang, Jia-Hui",
    title = "{Radiative properties and images of the rotating short-hairy black hole surrounded by a thin accretion disk}",
    eprint = "2501.10716",
    archivePrefix = "arXiv",
    primaryClass = "gr-qc",
    doi = "10.1103/mfxv-d98w",
    journal = "Phys. Rev. D",
    volume = "112",
    number = "6",
    pages = "064090",
    year = "2025"
}

@article{Tang:2025sbh,
    author = "Tang, Ruijing and Faraji, Shokoufe and Afshordi, Niayesh",
    title = "{Exploring Born-Infeld f(T) teleparallel gravity through accretion disk dynamics}",
    eprint = "2502.03775",
    archivePrefix = "arXiv",
    primaryClass = "gr-qc",
    doi = "10.1103/6q8h-k57n",
    journal = "Phys. Rev. D",
    volume = "113",
    number = "12",
    pages = "123021",
    year = "2026"
}

@article{Kumar:2025sxj,
    author = "Kumar, Dilip",
    title = "{Effect of noncommutative geometry on accretion disks around RGI-Schwarzschild black hole}",
    eprint = "2507.13056",
    archivePrefix = "arXiv",
    primaryClass = "gr-qc",
    reportNumber = "vol. 42(24), 245001",
    doi = "10.1088/1361-6382/ae255d",
    journal = "Class. Quant. Grav.",
    volume = "42",
    number = "24",
    pages = "245001",
    year = "2025"
}

@article{Boshkayev:2022vlv,
    author = "Boshkayev, Kuantay and Konysbayev, Talgar and Kurmanov, Yergali and Luongo, Orlando and Malafarina, Daniele",
    title = "{Accretion Disk Luminosity for Black Holes Surrounded by Dark Matter with Tangential Pressure}",
    eprint = "2205.04208",
    archivePrefix = "arXiv",
    primaryClass = "gr-qc",
    doi = "10.3847/1538-4357/ac8804",
    journal = "Astrophys. J.",
    volume = "936",
    number = "2",
    pages = "96",
    year = "2022"
}

@article{Kurmanov:2021uqv,
    author = "Kurmanov, Ergali and Boshkayev, Kuantay and Giamb{\`o}, Roberto and Konysbayev, Talgar and Luongo, Orlando and Malafarina, Daniele and Quevedo, Hernando",
    title = "{Accretion Disk Luminosity for Black Holes Surrounded by Dark Matter with Anisotropic Pressure}",
    eprint = "2110.15402",
    archivePrefix = "arXiv",
    primaryClass = "astro-ph.HE",
    doi = "10.3847/1538-4357/ac41d4",
    journal = "Astrophys. J.",
    volume = "925",
    number = "2",
    pages = "210",
    year = "2022"
}

@article{Heydari-Fard:2024wgu,
    author = "Heydari-Fard, Mohaddese and Heydari-Fard, Malihe and Riazi, Nematollah",
    title = "{Galactic black hole immersed in a dark halo with its surrounding thin accretion disk}",
    eprint = "2408.16020",
    archivePrefix = "arXiv",
    primaryClass = "gr-qc",
    doi = "10.1007/s10714-025-03382-5",
    journal = "Gen. Rel. Grav.",
    volume = "57",
    number = "2",
    pages = "49",
    year = "2025"
}

@article{Kobialko:2025sls,
    author = "Kobialko, Kirill and Gal'tsov, Dmitri and Molchanov, Alexey",
    title = "{Gravitational shadow and emission spectrum of thin accretion disks in a plasma medium}",
    eprint = "2505.07993",
    archivePrefix = "arXiv",
    primaryClass = "gr-qc",
    doi = "10.1103/lm4m-v7hj",
    journal = "Phys. Rev. D",
    volume = "112",
    number = "4",
    pages = "044039",
    year = "2025"
}

@article{Capolupo:2025dxx,
    author = "Capolupo, Antonio and Luongo, Orlando and Quaranta, Aniello",
    title = "{Impact of flavor condensate dark matter on accretion disk luminosity in spherical spacetimes}",
    eprint = "2507.03758",
    archivePrefix = "arXiv",
    primaryClass = "gr-qc",
    doi = "10.1140/epjc/s10052-026-15371-2",
    journal = "Eur. Phys. J. C",
    volume = "86",
    number = "2",
    pages = "154",
    year = "2026"
}

@article{Mummery:2024mrq,
    author = "Mummery, Andrew and Ingram, Adam and Davis, Shane and Fabian, Andrew",
    title = "{Continuum emission from within the plunging region of black hole discs}",
    eprint = "2405.09175",
    archivePrefix = "arXiv",
    primaryClass = "astro-ph.HE",
    doi = "10.1093/mnras/stae1160",
    journal = "Mon. Not. Roy. Astron. Soc.",
    volume = "531",
    number = "1",
    pages = "366--386",
    year = "2024"
}

@ARTICLE{2011ApJ...731..121B,
       author = {{Bambi}, Cosimo and {Barausse}, Enrico},
        title = "{Constraining the Quadrupole Moment of Stellar-mass Black Hole Candidates with the Continuum Fitting Method}",
      journal = {\apj},
         year = 2011,
        month = apr,
       volume = {731},
       number = {2},
          eid = {121},
        pages = {121},
          doi = {10.1088/0004-637X/731/2/121},
archivePrefix = {arXiv},
       eprint = {1012.2007},
 primaryClass = {gr-qc},
       adsurl = {https://ui.adsabs.harvard.edu/abs/2011ApJ...731..121B}
}

@article{Bambi:2015ldr,
    author = "Bambi, Cosimo and Jiang, Jiachen and Steiner, James F.",
    title = "{Testing the no-hair theorem with the continuum-fitting and the iron line methods: a short review}",
    eprint = "1511.07587",
    archivePrefix = "arXiv",
    primaryClass = "gr-qc",
    doi = "10.1088/0264-9381/33/6/064001",
    journal = "Class. Quant. Grav.",
    volume = "33",
    number = "6",
    pages = "064001",
    year = "2016"
}

@article{Luminet:1979nyg,
    author = "Luminet, J. -P.",
    title = "{Image of a spherical black hole with thin accretion disk}",
    journal = "Astron. Astrophys.",
    volume = "75",
    pages = "228--235",
    year = "1979"
}

@article{Zhang:2021hit,
    author = "Zhang, Zelin and Chen, Songbai and Qin, Xin and Jing, Jiliang",
    title = "{Polarized image of a Schwarzschild black hole with a thin accretion disk as photon couples to Weyl tensor}",
    eprint = "2106.07981",
    archivePrefix = "arXiv",
    primaryClass = "gr-qc",
    doi = "10.1140/epjc/s10052-021-09786-2",
    journal = "Eur. Phys. J. C",
    volume = "81",
    number = "11",
    pages = "991",
    year = "2021"
}

@article{Zhang:2022klr,
    author = "Zhang, Zelin and Chen, Songbai and Jing, Jiliang",
    title = "{Image of Bonnor black dihole with a thin accretion disk and its polarization information}",
    eprint = "2205.13696",
    archivePrefix = "arXiv",
    primaryClass = "gr-qc",
    doi = "10.1140/epjc/s10052-022-10794-z",
    journal = "Eur. Phys. J. C",
    volume = "82",
    number = "9",
    pages = "835",
    year = "2022"
}

@article{Hu:2023bzy,
    author = "Hu, Shiyang and Deng, Chen and Guo, Sen and Wu, Xin and Liang, Enwei",
    title = "{Observational signatures of Schwarzschild-MOG black holes in scalar{\textendash}tensor{\textendash}vector gravity: images of the accretion disk}",
    doi = "10.1140/epjc/s10052-023-11411-3",
    journal = "Eur. Phys. J. C",
    volume = "83",
    number = "3",
    pages = "264",
    year = "2023"
}

@article{Guo:2023zwy,
    author = "Guo, Sen and Huang, Yu-Xiang and Li, Guo-Ping",
    title = "{Optical appearance of the Schwarzschild black hole in the string cloud context*}",
    eprint = "2305.00007",
    archivePrefix = "arXiv",
    primaryClass = "gr-qc",
    doi = "10.1088/1674-1137/accad5",
    journal = "Chin. Phys. C",
    volume = "47",
    number = "6",
    pages = "065105",
    year = "2023"
}

@article{Huang:2023ilm,
    author = "Huang, Yu-Xiang and Guo, Sen and Cui, Yu-Hao and Jiang, Qing-Quan and Lin, Kai",
    title = "{Influence of accretion disk on the optical appearance of the Kazakov-Solodukhin black hole}",
    eprint = "2311.00302",
    archivePrefix = "arXiv",
    primaryClass = "gr-qc",
    doi = "10.1103/PhysRevD.107.123009",
    journal = "Phys. Rev. D",
    volume = "107",
    number = "12",
    pages = "123009",
    year = "2023"
}

@article{Cui:2024wvz,
    author = "Cui, Yu-Hao and Guo, Sen and Huang, Yu-Xiang and Liang, Yu and Lin, Kai",
    title = "{Optical appearance of numerical black hole solutions in higher derivative gravity}",
    eprint = "2408.03387",
    archivePrefix = "arXiv",
    primaryClass = "gr-qc",
    doi = "10.1140/epjc/s10052-024-13153-2",
    journal = "Eur. Phys. J. C",
    volume = "84",
    number = "8",
    pages = "772",
    year = "2024"
}

@article{Li:2024kyv,
    author = "Li, Zhen",
    title = "{Observational features of thin accretion disk around rotating regular black hole}",
    eprint = "2412.08447",
    archivePrefix = "arXiv",
    primaryClass = "gr-qc",
    doi = "10.1140/epjc/s10052-025-14255-1",
    journal = "Eur. Phys. J. C",
    volume = "85",
    number = "5",
    pages = "514",
    year = "2025"
}

@article{Li:2025ixk,
    author = "Li, Zhen and Guo, Xiao-Kan",
    title = "{Thin accretion disk around rotating hairy black hole: radiative property and optical appearance}",
    eprint = "2501.01018",
    archivePrefix = "arXiv",
    primaryClass = "gr-qc",
    doi = "10.1140/epjc/s10052-025-14423-3",
    journal = "Eur. Phys. J. C",
    volume = "85",
    number = "6",
    pages = "679",
    year = "2025"
}

@article{Yin:2025coq,
    author = "Yin, Jia-Jun and He, Tong-Yu and Liu, Ming and Fan, Hui-Min and Chen, Bohai and Han, Zhan-Wen and Yang, Rong-Jia",
    title = "{Observational properties of black hole in quantum fluctuation modified gravity}",
    eprint = "2503.00488",
    archivePrefix = "arXiv",
    primaryClass = "gr-qc",
    doi = "10.1016/j.nuclphysb.2025.117004",
    journal = "Nucl. Phys. B",
    volume = "1018",
    pages = "117004",
    year = "2025"
}

@article{Cai:2025rst,
    author = "Cai, Ziqiang and Ban, Zhenglong and Wang, Lu and Feng, Haiyuan and Long, Zheng-Wen",
    title = "{Thin accretion disk around Schwarzschild-like black hole in bumblebee gravity}",
    eprint = "2503.08424",
    archivePrefix = "arXiv",
    primaryClass = "gr-qc",
    doi = "10.1088/1475-7516/2025/10/101",
    journal = "\jcap",
    volume = "10",
    pages = "101",
    year = "2025"
}

@article{Cai:2025pan,
    author = "Cai, Ziqiang and Ban, Zhenglong and Wang, Lu and Feng, Haiyuan and Long, Zheng-Wen",
    title = "{Shadow and thin accretion disk around Ay{\'o}n-Beato{\textendash}Garc{\'\i}a black hole coupled with cloud of strings}",
    eprint = "2506.22744",
    archivePrefix = "arXiv",
    primaryClass = "gr-qc",
    doi = "10.1016/j.dark.2025.102169",
    journal = "Phys. Dark Univ.",
    volume = "50",
    pages = "102169",
    year = "2025"
}

@article{Gyulchev:2026kvp,
    author = "Gyulchev, Galin N. and Doneva, Daniela D. and Deliyski, Valentin O. and Nedkova, Petya G. and Yazadjiev, Stoytcho S.",
    title = "{Images of the thin accretion disk around Kerr black holes coupled to time periodic scalar fields}",
    eprint = "2603.08796",
    archivePrefix = "arXiv",
    primaryClass = "gr-qc",
    doi = "10.1103/3xwd-4p4w",
    journal = "Phys. Rev. D",
    volume = "114",
    number = "2",
    pages = "024074",
    year = "2026"
}

@ARTICLE{2005ApJS..157..335L,
       author = {{Li}, Li-Xin and {Zimmerman}, Erik R. and {Narayan}, Ramesh and {McClintock}, Jeffrey E.},
        title = "{Multitemperature Blackbody Spectrum of a Thin Accretion Disk around a Kerr Black Hole: Model Computations and Comparison with Observations}",
      journal = {\apjs},
         year = 2005,
        month = apr,
       volume = {157},
       number = {2},
        pages = {335-370},
          doi = {10.1086/428089},
archivePrefix = {arXiv},
       eprint = {astro-ph/0411583},
 primaryClass = {astro-ph},
       adsurl = {https://ui.adsabs.harvard.edu/abs/2005ApJS..157..335L}
}

@ARTICLE{2011ApJ...742...85G,
       author = {{Gou}, Lijun and {McClintock}, Jeffrey E. and {Reid}, Mark J. and {Orosz}, Jerome A. and {Steiner}, James F. and {Narayan}, Ramesh and {Xiang}, Jingen and {Remillard}, Ronald A. and {Arnaud}, Keith A. and {Davis}, Shane W.},
        title = "{The Extreme Spin of the Black Hole in Cygnus X-1}",
      journal = {\apj},
         year = 2011,
        month = dec,
       volume = {742},
       number = {2},
          eid = {85},
        pages = {85},
          doi = {10.1088/0004-637X/742/2/85},
archivePrefix = {arXiv},
       eprint = {1106.3690},
 primaryClass = {astro-ph.HE},
       adsurl = {https://ui.adsabs.harvard.edu/abs/2011ApJ...742...85G}
}

@ARTICLE{2014ApJ...790...29G,
       author = {{Gou}, Lijun and {McClintock}, Jeffrey E. and {Remillard}, Ronald A. and {Steiner}, James F. and {Reid}, Mark J. and {Orosz}, Jerome A. and {Narayan}, Ramesh and {Hanke}, Manfred and {Garc{\'\i}a}, Javier},
        title = "{Confirmation via the Continuum-fitting Method that the Spin of the Black Hole in Cygnus X-1 Is Extreme}",
      journal = {\apj},
         year = 2014,
        month = jul,
       volume = {790},
       number = {1},
          eid = {29},
        pages = {29},
          doi = {10.1088/0004-637X/790/1/29},
archivePrefix = {arXiv},
       eprint = {1308.4760},
 primaryClass = {astro-ph.HE},
       adsurl = {https://ui.adsabs.harvard.edu/abs/2014ApJ...790...29G}
}

@article{Moore:2015bxa,
    author = "Moore, Christopher J. and Gair, Jonathan R.",
    title = "{Testing the no-hair property of black holes with x-ray observations of accretion disks}",
    eprint = "1507.02998",
    archivePrefix = "arXiv",
    primaryClass = "gr-qc",
    doi = "10.1103/PhysRevD.92.024039",
    journal = "Phys. Rev. D",
    volume = "92",
    number = "2",
    pages = "024039",
    year = "2015"
}

@article{Tripathi:2021rqs,
    author = "Tripathi, Ashutosh and Abdikamalov, Askar B. and Ayzenberg, Dimitry and Bambi, Cosimo and Grinberg, Victoria and Liu, Honghui and Zhou, Menglei",
    title = "{Testing the Kerr black hole hypothesis with the continuum-fitting and the iron line methods: the case of GRS~1915+105}",
    eprint = "2106.10982",
    archivePrefix = "arXiv",
    primaryClass = "astro-ph.HE",
    doi = "10.1088/1475-7516/2022/01/019",
    journal = "\jcap",
    volume = "01",
    number = "01",
    pages = "019",
    year = "2022"
}

@article{Zhang:2021ymo,
    author = "Zhang, Zuobin and Liu, Honghui and Abdikamalov, Askar B. and Ayzenberg, Dimitry and Bambi, Cosimo and Zhou, Menglei",
    title = "{Testing the Kerr Black Hole Hypothesis with GRS 1716-249 by Combining the Continuum Fitting and the Iron-line Methods}",
    eprint = "2106.03086",
    archivePrefix = "arXiv",
    primaryClass = "astro-ph.HE",
    doi = "10.3847/1538-4357/ac350e",
    journal = "Astrophys. J.",
    volume = "924",
    number = "2",
    pages = "72",
    year = "2022"
}

@article{Zhao:2021bhj,
    author = "Zhao, Xueshan and Gou, Lijun and Dong, Yanting and Zheng, Xueying and Steiner, James F. and Miller-Jones, James C. A. and Bahramian, Arash and Orosz, Jerome A. and Feng, Ye",
    title = "{Re-estimating the Spin Parameter of the Black Hole in Cygnus X-1}",
    eprint = "2102.09093",
    archivePrefix = "arXiv",
    primaryClass = "astro-ph.HE",
    doi = "10.3847/1538-4357/abbcd6",
    journal = "Astrophys. J.",
    volume = "908",
    number = "2",
    pages = "117",
    year = "2021"
}

@article{Drew:2025euq,
    author = "Drew, Michael and Stanway, Joshua S. and Patterson, Brett A. and Walton, Timothy J. and Ward-Thompson, Derek",
    title = "{New Estimates of the Spin and Accretion Rate of the Black Hole M87*}",
    eprint = "2505.17035",
    archivePrefix = "arXiv",
    primaryClass = "astro-ph.HE",
    doi = "10.3847/2041-8213/adc90e",
    journal = "Astrophys. J. Lett.",
    volume = "984",
    number = "1",
    pages = "L31",
    year = "2025"
}

@article{EventHorizonTelescope:2022urf,
    author = "Akiyama, Kazunori and others",
    collaboration = "Event Horizon Telescope",
    title = "{First Sagittarius A* Event Horizon Telescope Results. V. Testing Astrophysical Models of the Galactic Center Black Hole}",
    eprint = "2311.09478",
    archivePrefix = "arXiv",
    primaryClass = "astro-ph.HE",
    reportNumber = "FERMILAB-PUB-22-419-PPD",
    doi = "10.3847/2041-8213/ac6672",
    journal = "Astrophys. J. Lett.",
    volume = "930",
    number = "2",
    pages = "L16",
    year = "2022"
}
\end{document}